\documentclass{ws-ijmpe}
\usepackage[super,compress]{cite}
\usepackage[spaces,hyphens]{url} 
\newcommand{\dif}{\mathrm{d}}
\begin{document}

\markboth{P.~Huovinen}{Hydrodynamics at RHIC and LHC: What have we learned?}

\catchline{}{}{}{}{}

\title{Hydrodynamics at RHIC and LHC: What have we learned?}

\author{\footnotesize PASI HUOVINEN}

\address{Frankfurt Institute for Advanced Studies, Ruth-Moufang-Stra\ss e 1\\
60438 Frankfurt am Main, Germany\\
huovinen@fias.uni-frankfurt.de}

\maketitle

\begin{history}
\end{history}

\begin{abstract}
Fluid dynamical description of elementary particle collisions has a
long history dating back to the works of Landau and Fermi.
Nevertheless, it is during the last 10--15 years when fluid dynamics
has become the standard tool to describe the evolution of matter
created in ultrarelativistic heavy ion collisions. In this review I
briefly describe the hydrodynamical models, what we have learned when
analyzing the RHIC and LHC data using these models, and what the
latest developments and challenges are.
\end{abstract}


\ccode{PACS numbers: 25.75Ld, 25.75.Nq, 24.10Nz}


\section{Introduction}

The goal of the heavy-ion programs at BNL RHIC\footnote{Relativistic
  Heavy Ion Collider, full collision energy 
  $\sqrt{s_\mathrm{NN}} = 200$ GeV.} and CERN LHC\footnote{Large Hadron
  Collider, $\sqrt{s_\mathrm{NN}} = 2.76$ TeV at the time of this
  writing, designed $\sqrt{s_\mathrm{NN}} = 5.5$ TeV.} is to observe
strongly interacting matter. 'Strongly interacting' in a sense that
the interactions in the system are not mediated by the
electromagnetic, but by the strong interaction, and 'matter' in the
sense that to describe the system, we do not need to describe every
(quasi-)particle individually, but thermodynamic concepts like
temperature and pressure are applicable. Thus it is natural to try to
describe the expansion stage of the collision by a macroscopic
approach like fluid dynamics.  We expect that the system formed in the
collision is initially so hot and dense that relevant degrees of
freedom are partons, not hadrons. The system expands and cools,
undergoes a phase transition to hadrons, and when the system is dilute
enough, interactions cease and the particles stream freely to
detectors.

\section{Fluid dynamics}

\subsection{Ideal fluid dynamics}
 \label{ideal}

Relativistic fluid dynamics is basically an application of
conservation laws for energy, momentum and conserved charges (if
any). When written in differential form,
\begin{equation}
 \partial_\mu T^{\mu\nu} = 0, \qquad\mathrm{and}\qquad \partial_\mu j_x^\mu = 0,
\end{equation}
where $T^{\mu\nu}$ is the energy-momentum tensor, and $j_x^\mu$ the
charge 4-current of charge $x$ (baryon or electric charge,
strangeness, isospin,\ldots), the conservation laws provide evolution
equations for the system. If there are $n$ conserved charges, there
are $4+n$ equations, which contain $10+4n$ unknowns. To make this set
of equations solvable, further constraints for the unknowns must be
provided. The simplest approach is to assume that the system is in
exact local equilibrium. In that case the energy-momentum tensor and
charge currents can be expressed as
\begin{equation}
 T^{\mu\nu} = (\epsilon+P)u^\mu u^\nu - P g^{\mu\nu},
  \qquad\mathrm{and}\qquad
 j_x^\mu = n_x u^\mu,
\end{equation}
respectively, where $\epsilon$ and $P$ are energy density and pressure
in the local rest frame, $u^\mu$ flow velocity, $g^{\mu\nu}$ is the
metric tensor, and $n_x$ the charge density of charge $x$ in the local
rest frame. Now the number of unknowns is reduced to $5+n$, and the
system of equations can be closed by providing the equation of state
(EoS) of the fluid in a form $P = P(\epsilon,\{n_i\})$. To model the
collisions at RHIC and LHC, the description is usually further
simplified by assuming that net baryon density and other conserved
charges are zero.  After specifying the boundary conditions for this
set of partial differential equations the evolution is determined, and
all the microscopic physics is contained within the EoS\footnote{For
  further discussion of fluid dynamics see
  Refs.~\refcite{Rischke:1998fq} and~\refcite{Ollitrault:2007du}.}.

However, the simple phrase ``after specifying the boundary
conditions'' contains a lot of physics. Hydrodynamics provides neither
the initial distribution of matter nor the criterion for the end of
the evolution, but they must be supplied by other models. At
ultrarelativistic energies the initial state of the hydrodynamical
evolution cannot be two colliding nuclei: The initial collision
processes are far from equilibrium and produce large amount of
entropy~\cite{Muller:2011ra}. Thus the initial state is assumed to be
a distribution of thermalized matter soon after the initial
collision. The most common approach is so-called Glauber
model~\cite{Miller:2007ri,Bialas:1976ed}, which is basically a
geometrical constraint on the distributions. The Woods-Saxon
distributions of nuclear matter in colliding nuclei are projected on a
plane orthogonal to the beam (so called transverse plane), and the
resulting densities on this plane, and nucleon-nucleon cross section
at the collision energy, are used to calculate the number density of
binary collisions and participants on this plane---participant meaning
a nucleon which has interacted at least once (for details, see
Ref.~\refcite{Miller:2007ri}). The initial energy or entropy density
profile is taken to be proportional to the profile of collisions or
participants, or to a linear combination of
them~\cite{Kolb:2001qz,Heinz:2001xi}. The proportionality constant is
a free parameter chosen to reproduce the observed final particle
multiplicity.

Another popular approach is the KLN (Kharzeev-Levin-Nardi)
model~\cite{Kharzeev:2000ph,Kharzeev:2001gp,Kharzeev:2001yq,Kharzeev:2002ei,Drescher:2006ca,Drescher:2007ax}
where the initial entropy density distribution is proportional to the
distribution of gluons produced in primary collisions. The gluon
production is calculated using the Color Glass Condensate (CGC)
framework~\cite{Iancu:2002xk,Iancu:2003xm,Gelis:2010nm,Lappi:2010ek},
where one applies the feature of QCD that at small-x gluon densities
are large. These large densities correspond to classical fields
permitting calculations using classical techniques. Another approach
to calculate the initial particle production from first principles is
so-called EKRT saturation
model~\cite{Eskola:1999fc,Paatelainen:2012at,Paatelainen:2013eea}
based on perturbative QCD + saturation framework. Besides these
approaches, one can use event generators like
UrQMD~\cite{Bass:1998ca,Bleicher:1999xi,Petersen:2008dd},
AMPT~\cite{Zhang:1999bd,Pang:2012he}, or EPOS~\cite{Werner:2010aa} to
generate the initial state for hydrodynamic expansion. Nevertheless,
no model so far describes a dynamical process leading to thermalized
matter, but thermalization has to be postulated and imposed by hand.

When the system expands and cools, the mean free paths
increase. Ultimately mean free paths become so long that rescatterings
cease, and particle distributions no longer evolve. Particle
distributions are 'frozen' at that stage, and we say that the
particles freeze out, or alternatively, that the particles decouple
from each other. In practice this freeze-out should be a gradual
process, but since implementing a gradual freeze-out is
difficult\footnote{For attempts see
  Refs.~\refcite{Akkelin:2008eh,Karpenko:2010te}}, it is approximated
to take place suddenly on a surface of zero thickness. In such a case
one can use so-called Cooper-Frye prescription to evaluate particle
distributions on this surface~\cite{Cooper:1974mv}:
\begin{equation}
 \label{C-F}
 E\frac{\dif N}{\dif p^3} = \int_\sigma \dif\sigma_\mu p^\mu 
                            f(T(x),\mu (x),p\cdot u(x)),
\end{equation}
where $\sigma$ is the surface where the distribution is to be
evaluated, $\dif\sigma_\mu$ its normal 4--vector, $f$ distribution of
particles, and $u$ 4--flow velocity of the fluid. Hydrodynamics does
not tell when decoupling should take place, but the criterion is a
free parameter, and its value is chosen to reproduce the observed
$p_T$-spectra. Usually a constant temperature or energy density is
used as the criterion, although a more realistic criterion would be
the ratio of the scattering rate of particles to the expansion rate of
the system, \emph{i.e.}, the inverse Knudsen
number~\cite{Bondorf:1978kz,Holopainen:2012id}

\subsection{Dissipative fluid dynamics}

The ideal fluid assumption is extremely strong, and in nature
gradients in the system always indicate deviations from equilibrium
and thus dissipation. In non-relativistic fluid dynamics Navier-Stokes
equations are known to describe viscous fluid well, but unfortunately
the relativistic generalization of Navier-Stokes equations is unstable
and allows acausal solutions with superluminal signal propagation
speeds~\cite{Hiscock1,Hiscock2,Hiscock3}. This undesired behavior can
be avoided if one assumes that the dissipative currents (shear stress
$\pi^{\mu\nu}$, heat flow $q^\mu$ and bulk pressure $\Pi$) are not
directly proportional to gradients in the system, but are dynamical
variables which relax to their Navier-Stokes values on time scales
given by corresponding relaxation times $\tau_\pi$, $\tau_q$ and
$\tau_\Pi$.  The evolution equations for the dissipative currents can
be derived phenomenologically from entropy
current~\cite{Muller:1967,Israel}, from kinetic theory using the Grad
14--moment ansatz~\cite{Grad,IS,Denicol:2012es}, or via gradient
expansion~\cite{Baier:2007ix,Romatschke:2009kr}.

The derivation from kinetic theory leads to the commonly used
Israel-Stewart equations, but so far it has had a problem: Unlike
Chapman-Enskog expansion~\cite{Chapman}, which leads to relativistic
Navier-Stokes equations, it is not a controlled expansion in some
small parameter, in which one could do power counting and improve the
approximation if necessary. This problem has recently been solved by
rederiving viscous hydrodynamics using the method of moments, and
ordering the terms in the expansion according to their (generalized)
Reynolds and Knudsen numbers~\cite{Denicol:2012cn}. In lowest order in
Knudsen and Reynolds numbers this approach leads to equations
identical to the original Israel-Stewart approach, but it has been
shown that an adequate description of heat flow would require
inclusion of some higher order terms\cite{Denicol:2012vq}.

Strictly speaking all these derivations are for a single component
system, and the derivation for a multi-component system has not been
completed yet~\cite{Prakash:1993bt,Monnai:2010qp,Denicol:2012yr}. For
the behavior of the fluid this does not matter---the different
components are assumed to behave as a single fluid---but when the
fluid is converted to various types of hadrons, this causes an
additional uncertainty. The Cooper-Frye prescription (Eq.~(\ref{C-F}))
is equally valid in dissipative as in ideal system, but dissipation
causes the distribution function $f$ to deviate slightly from the
thermal equilibrium distribution. If there is only shear, no bulk
pressure nor heat flow, in a single component system the Grad
14--moment ansatz leads in Boltzmann approximation to
\begin{equation}
  \label{delta-f}
 f = f_0 + \delta f, \qquad\mathrm{where}\qquad 
 \delta f = f_0\frac{p^\mu p^\nu \pi_{\mu\nu}}{2 T^2(\epsilon+P)},
\end{equation}
and $f_0$ is the equilibrium distribution function. In a
multicomponent system the effect of shear stress on particle
distributions should depend on particle properties, but how exactly,
is still a work in progress~\cite{Molnar:2011kx}. In practical
calculations one has therefore assumed that the dissipative correction
to the thermal distribution is given by Eq.~(\ref{delta-f}) for all
hadron species. Note that the form of $\delta f$ in
Eq.~(\ref{delta-f}) is only an ansatz. Other forms have been argued
for~\cite{Dusling:2009df}, but if the thermal distribution is expanded
differently than in the 14--moment ansatz, the evolution equations and
definitions of the dissipative coefficients may change as well.

So far the viscous hydrodynamical calculations have concentrated on
studying the effects of shear viscosity, characterized by the shear
viscosity coefficient $\eta$. In no calculation has heat conductivity
been included, and the bulk pressure has usually been omitted as
well\footnote{There are some calculations with bulk, like
  Refs.~\refcite{Monnai:2009ad,Song:2009rh,Bozek:2011ua,Dusling:2011fd,Noronha-Hostler:2013gga}.}.
When the only dissipative current is shear stress, the energy-momentum
tensor becomes
\begin{equation}
  T^{\mu\nu} = (\epsilon + P)u^\mu u^\nu - Pg^{\mu\nu} + \pi^{\mu\nu},
\end{equation}
where $\pi^{\mu\nu}$ is the shear stress tensor. Its evolution is
usually given by
\begin{equation}
 \label{pi-evo}
 \langle D\pi^{\mu\nu}\rangle = 
  \frac{1}{\tau_\pi} \left(2 \eta\sigma^{\mu \nu} - \pi^{\mu\nu} \right)
  -\frac{4}{3}\pi^{\mu\nu}\partial_\lambda u^\lambda,
\end{equation}
where $D=u^\mu \partial_\mu$, 
$\sigma^{\mu \nu} = \nabla^{\langle\mu} u^{\nu\rangle}$, and the
angular brackets $\langle\,\rangle$ denote the symmetrized and
traceless projection, orthogonal to the fluid four-velocity
$u^\mu$. Both Israel-Stewart~\cite{Denicol:2012es,Denicol:2012cn} and
gradient expansion~\cite{Baier:2007ix,Romatschke:2009kr} approaches
lead to some additional terms in the evolution equation, but these
terms are usually considered numerically insignificant, although a
full analysis of the effect of all terms has not yet been done. It is
worth noting that the factor 4/3 in the last term of
Eq.~(\ref{pi-evo}) is strictly valid for massless particles. Whether
modifying this term according to the mass of particles would change
any results has not been studied yet. As well, in the original
Israel-Stewart papers~\cite{Israel,IS} this term was estimated small
and omitted in the final equations. In the context of heavy-ion
collisions, however, it improves the applicability of viscous
hydrodynamics significantly~\cite{Huovinen:2008te}.

\subsection{Hybrid models}
 \label{hybrid}

The sudden change from interacting fluid to free streaming particles
in the Cooper-Frye description is clearly an oversimplification.  We
also expect hadron gas to be so
dissipative~\cite{Prakash:1993bt,Demir:2008tr} that the applicability
of dissipative fluid dynamics is questionable~\cite{Song:2010aq}. In
so-called hybrid models these problems are avoided by using fluid
dynamics to describe only the early dense stage of the evolution. The
fluid is converted to particles using the Cooper-Frye description, not
at freeze-out, but when rescatterings are still abundant, and the
individual particles are fed to a hadron cascade model like
UrQMD~\cite{Bass:1998ca,Bleicher:1999xi} or
JAM~\cite{Nara:1999dz,Isse:2005nk} to describe the dilute hadronic
stage. These models have the advantage that hadron cascades describe
freeze-out dynamically without free parameters, all dissipative
processes are included, and they can describe a system arbitrarily far
from equilibrium. Unfortunately they do not remove all the
arbitrariness from the end of the evolution: Like the results of pure
hydrodynamical models depend on the freeze-out criterion, the results
of hybrid models depend on when the switch from fluid to cascade is
done~\cite{Song:2010aq,Hirano:2012kj}. For a detailed discussion of
these models see Ref.~\refcite{Hirano:2012kj}.

\section{What we know}

\subsection{There are rescatterings}
 \label{thermal}

\begin{figure}
 \begin{minipage}[t]{0.48\textwidth}
   \includegraphics[width=\textwidth]{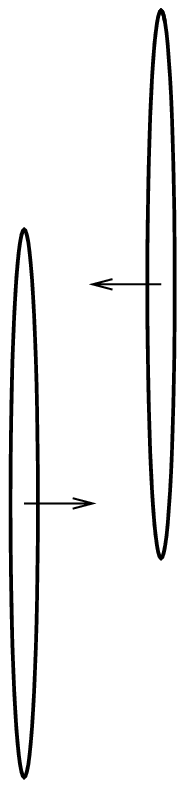}
 \end{minipage}
  \hfill
 \begin{minipage}[t]{0.48\textwidth}
  \includegraphics[width=\textwidth]{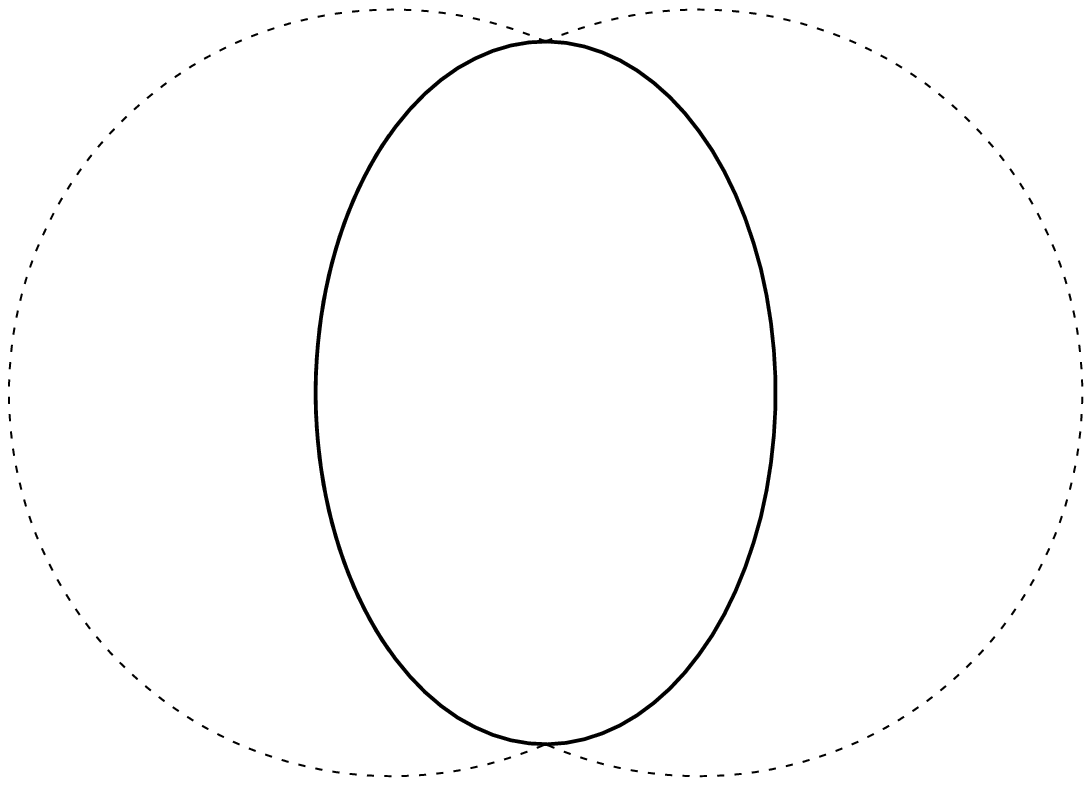}
 \end{minipage} 
 \caption{A schematic representation of the geometry of a non-central
   heavy-ion collision. (left:) Two Lorentz-contracted nuclei
   approaching each other. (right:) A projection of the collision zone
   onto a plane orthogonal to the beam (so called transverse plane).
   The dotted circles depict the target and projectile nuclei.}
 \label{schematic}
\end{figure}

The particle production in the primary collisions is azimuthally
isotropic, but the distribution of observed particles in A+A
collisions is not. The anisotropy can be easily explained in terms of
rescatterings of the produced particles: In a non-central collision of
identical nuclei the collision zone has an elongated shape, see
Fig.~\ref{schematic}. If a particle is moving along the long axis of
the collision zone, it has a larger probability to scatter and change
its direction than a particle moving along the short axis.  Thus more
particles end up in the direction of the short axis.  Or, in the
hydrodynamical language, the pressure gradient between the center of
the system and the vacuum is larger in the direction of the short
axis, the flow velocity is thus larger in that direction, and more
particles are emitted along the short than along the long axis.

\begin{figure}
 \begin{minipage}[t]{0.48\textwidth}
  \includegraphics[width=0.8\textwidth]{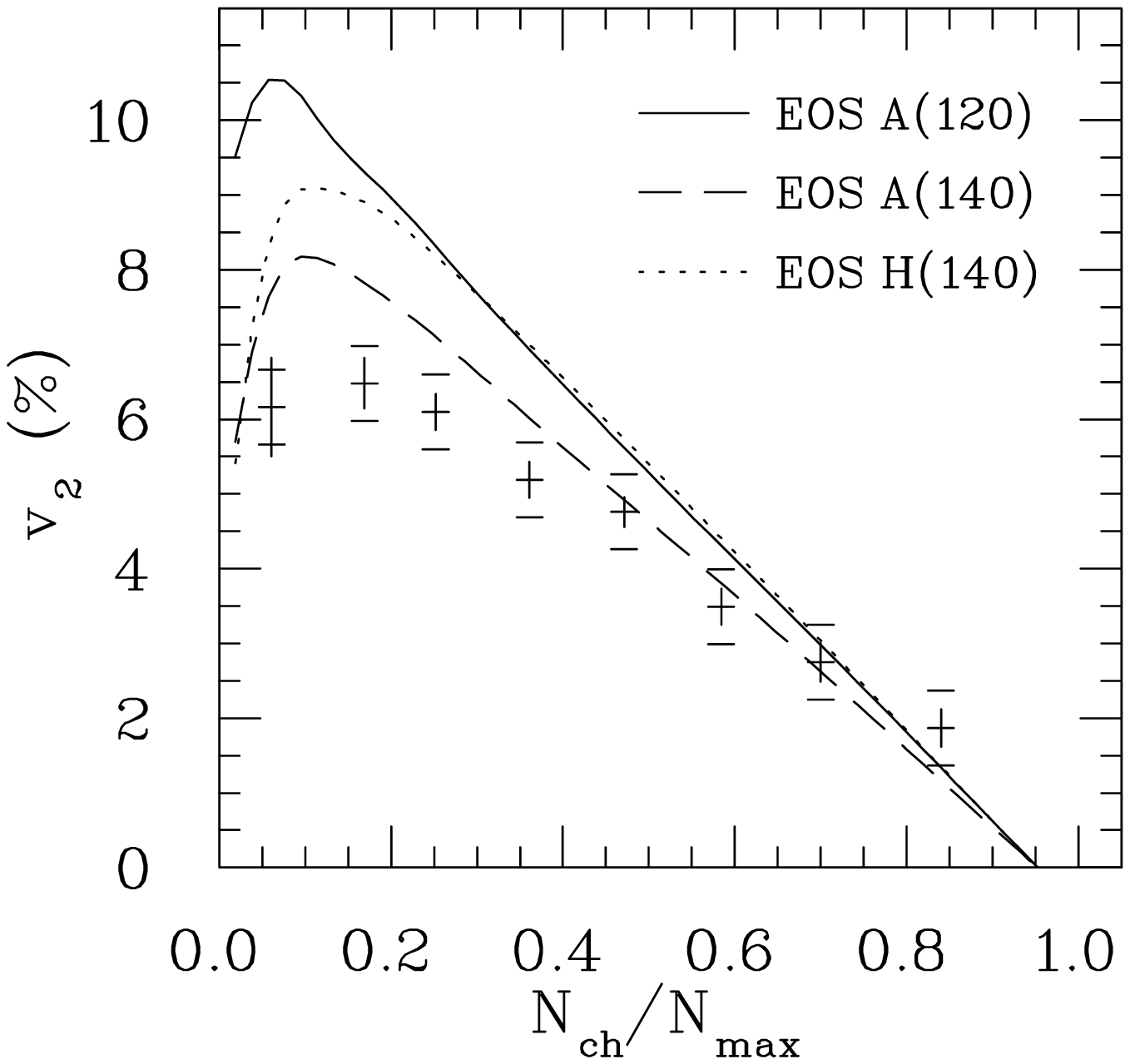}
  \caption{The elliptic flow parameter $v_2$ of charged hadrons as
    function of centrality ($\mathrm{N_{ch}/N_{max}}=1$ is the most
    central collision) in Au+Au collisions at
    $\sqrt{s_\mathrm{NN}}=130$ GeV. The data are from
    Ref.~\protect\refcite{Ackermann:2000tr} and the calculation using
    different EoSs (labels A and H), and freeze-out temperatures (120
    MeV or 140 MeV) from Ref.~\protect\refcite{Kolb:2000fha}. The figure
    is from Ref.~\protect\refcite{confinement}.}
  \label{ur-v2}
 \end{minipage}
  \hfill
 \begin{minipage}[t]{0.48\textwidth}
  \hspace*{-10mm}
  \includegraphics[width=1.2\textwidth]{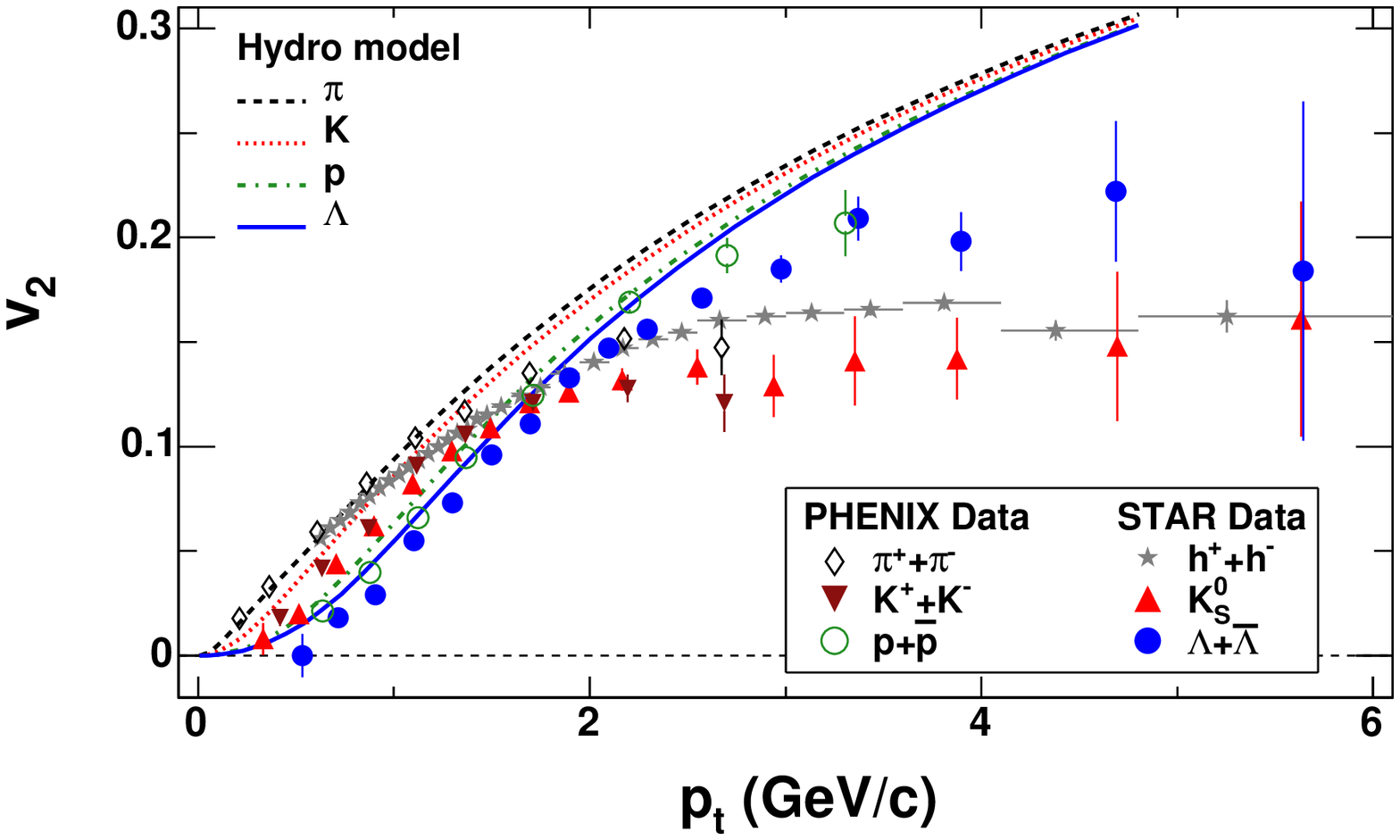}
  \caption{$v_2$ vs.~$p_T$ for identified particles in minimum bias
    $Au+Au$ collisions at $\sqrt{s_\mathrm{NN}}=200$ GeV. The data are
    from Refs.~\protect\refcite{Adams:2004bi,Adler:2003kt}
    and~\protect\refcite{Adams:2003am}. Reprinted figure with
    permission from Ref.~\protect\refcite{Adams:2004bi}. Copyright
    (2005) by the American Physical Society.}
  \label{massorder}
 \end{minipage}
\end{figure}

This anisotropy is quantified in terms of Fourier expansion of the
azimuthal distribution. The coefficients of this expansion $v_n$, and
the associated participant angles $\psi_n$, are defined
as\footnote{For a detailed discussion of the anisotropy measurements
  see Ref.~\refcite{Voloshin:2008dg}.}
\begin{equation}
v_n = \langle \cos[n(\phi-\psi_n)]\rangle,
\qquad\mathrm{and}\qquad
\psi_n = \frac{1}{n}\arctan\frac{\langle p_T\sin(n\phi)\rangle}
                                {\langle p_T\cos(n\phi)\rangle}.
\end{equation}
Of these coefficients $v_1$ is called directed, and $v_2$ elliptic
flow. Elliptic flow of charged hadrons as a function of centrality
was one of the first measurements at RHIC~\cite{Ackermann:2000tr}. The
result is shown in Fig.~\ref{ur-v2}, and compared to early fluid
dynamical calculations~\cite{Kolb:2000fha}. As seen, the elliptic flow
is quite large and increases with decreasing centrality, as expected
if it has the described geometric origin. Thus there must be
rescatterings among the particles formed in the collision, and an A+A
collision is not just a sum of independent $pp$ collisions.

The observed elliptic flow is also very close to the hydrodynamically
calculated one, which is a strong indication of hydrodynamical
behavior of the matter. Another signature of hydrodynamical behavior
is shown in Fig.~\ref{massorder}: It was observed that the heavier the
particle, the smaller its $p_T$-differential $v_2$ at low $p_T$. As
explained in Refs.~\refcite{Ollitrault:2007du}
and~\refcite{Huovinen:2001cy}, such a behavior arises if all the
particles are emitted from the same expanding thermal source. Thus, if
the produced matter is not close to kinetic equilibrium, at least it
behaves as if it was!

It has often been argued that the large observed $v_2$ and its
hydrodynamical reproduction requires that the system reaches thermal
equilibrium very fast~\cite{Kolb:2003dz}, within 1 fm/$c$ after the
initial collision. The hydrodynamical models fitting the data do
indeed use short initial times, $\tau_0 = 0.15$-1 fm/$c$, but in
Ref.~\refcite{Luzum:2008cw} it was shown that $\tau_0 = 2$ fm/$c$
works as well. The crucial distinction is what the shape of the
initial state is: The larger the deformation, the larger the final
momentum anisotropy. During thermalization the matter will not stay
put, but begins to expand. This expansion reduces the spatial
anisotropy, and unless momentum anisotropy is built up during
thermalization, the final anisotropy is smaller~\cite{Kolb:2000sd}. If
thermalization is fast, we may assume that changes in matter
distribution and flow field during thermalization are tiny, and
geometry (Glauber) and/or initial gluon production (Color Glass, EKRT)
give reasonable constraints to the initial state of hydrodynamical
evolution. In Ref.~\refcite{Luzum:2008cw} the shape of the initial
state was assumed to be independent of the thermalization time, but if
thermalization time is long, this is no longer a good
approximation. Thus, if thermalization is fast, we know that we can
reproduce the data using hydrodynamical models, but if thermalization
takes long, we do not know, since we do not know a plausible initial
state for hydrodynamical evolution.

\subsection{Equation of State has many degrees of freedom}

\begin{figure}
 \begin{minipage}[t]{0.49\textwidth}
  \includegraphics[width=\textwidth]{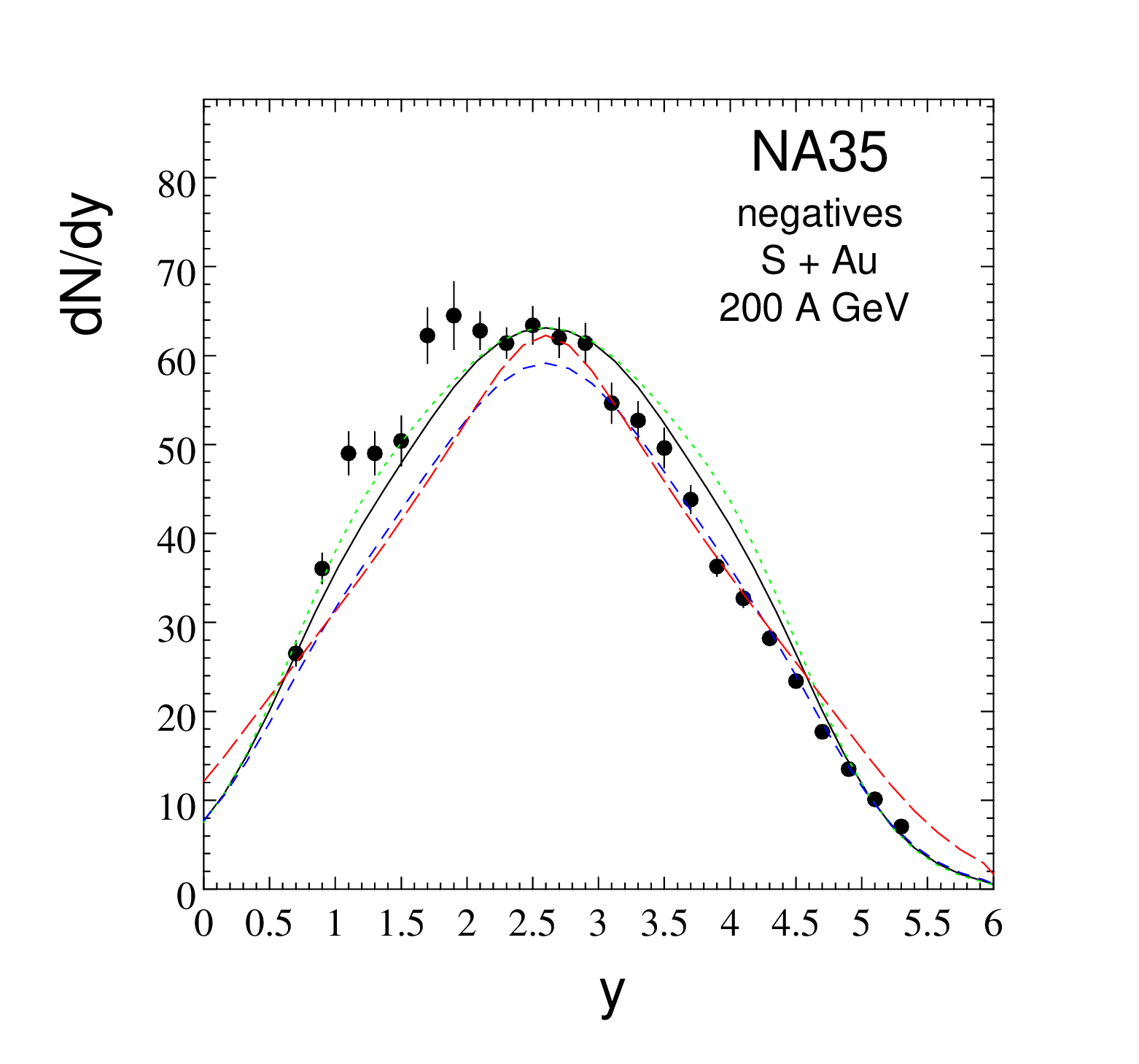}
 \end{minipage}
  \hfill
 \begin{minipage}[t]{0.49\textwidth}
  \includegraphics[width=\textwidth]{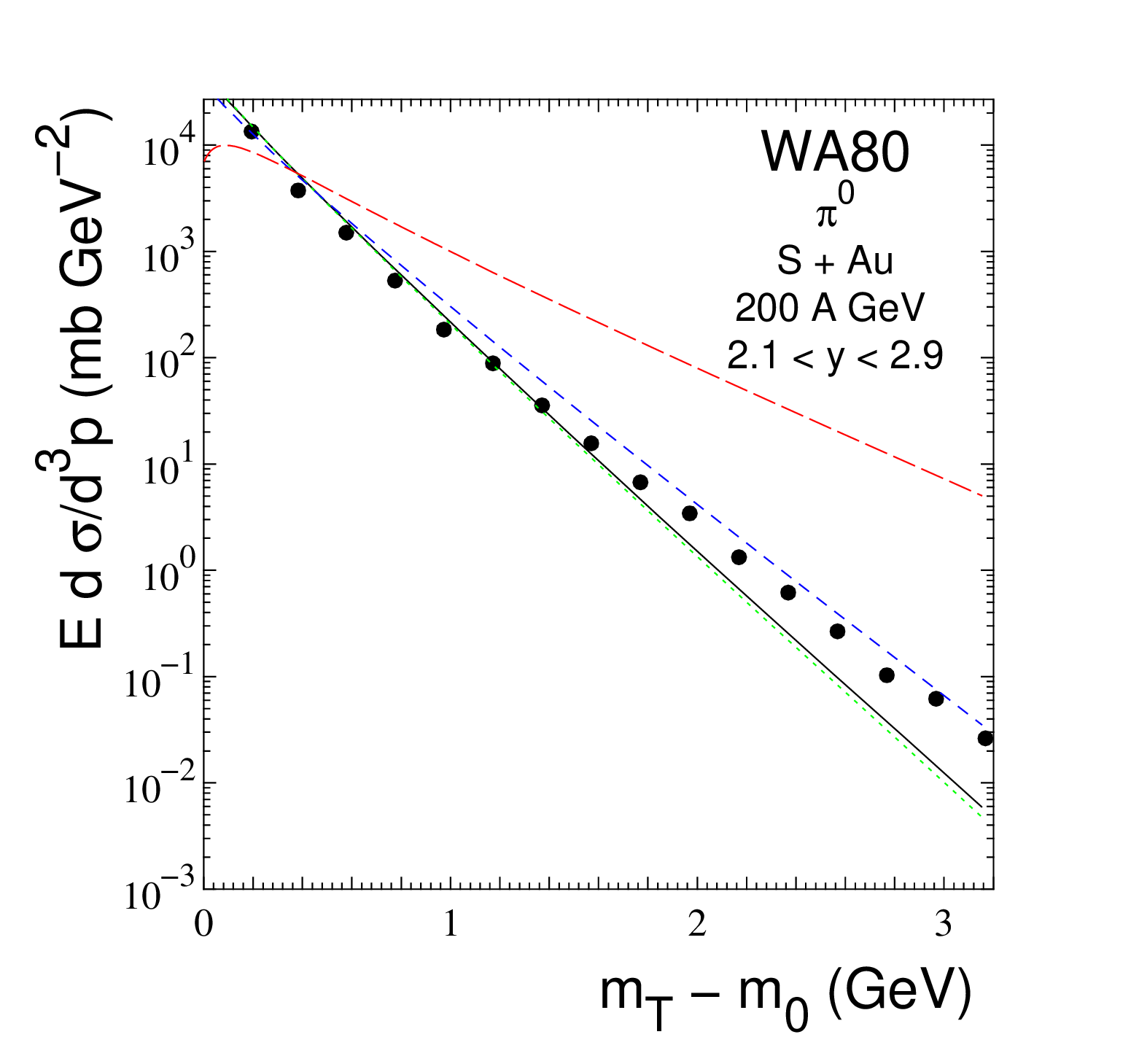}
 \end{minipage}
 \caption{Rapidity distribution of negative hadrons (left) and
   transverse momentum distribution of neutral pions (right) in S+Au
   collision at $E_{\mathrm{lab}} = 200A$ GeV using four different
   EoSs. The red long dashed curve corresponds to ideal pion gas EoS,
   whereas the other curves correspond to hadron resonance gas with or
   without phase transition to ideal parton gas. The figures are from
   Ref.~\protect\refcite{Sollfrank:1996hd}, and the data from
   Refs.~\protect\refcite{Gazdzicki:1995cr}
   and~\protect\refcite{Santo:1993ri}.}
  \label{idealpion}
\end{figure}

The equation of state (EoS) of strongly interacting matter is an
explicit input to hydrodynamical models. Thus one might expect
hydrodynamical modeling of heavy-ion collisions to tell us a lot about
the EoS, but unfortunately that is not the case. The collective motion
of the system is directly affected by the pressure gradients in the
system, and thus by the EoS, but the effects of the EoS on the final
particle $p_T$ distributions can to very large extent be compensated
by changes in the initial state of the evolution, and the final
decoupling temperature. This makes constraining the properties of the
EoS very difficult. However, what we do know is that the number of
degrees of freedom has to be large.

It was already seen when modeling S+Au collisions at the CERN SPS at
$E_\mathrm{lab} = 200A$ GeV energy, that if we use ideal pion gas EoS,
it is not possible to fit the pion rapidity and $p_T$-distributions
simultaneously. Once the initial state and decoupling temperature are
fixed to reproduce the rapidity distribution, the transverse momentum
distribution of pions becomes too flat~\cite{Sollfrank:1996hd}, see
Fig.~\ref{idealpion}. Or if one chooses parameters to fit the
$p_T$-spectrum, the rapidity distribution is not reproduced.  On the
other hand, if we use an EoS containing several hadrons and
resonances, the distributions can be fitted.

\begin{figure}
 \begin{center}
   \includegraphics[width=0.49\textwidth]{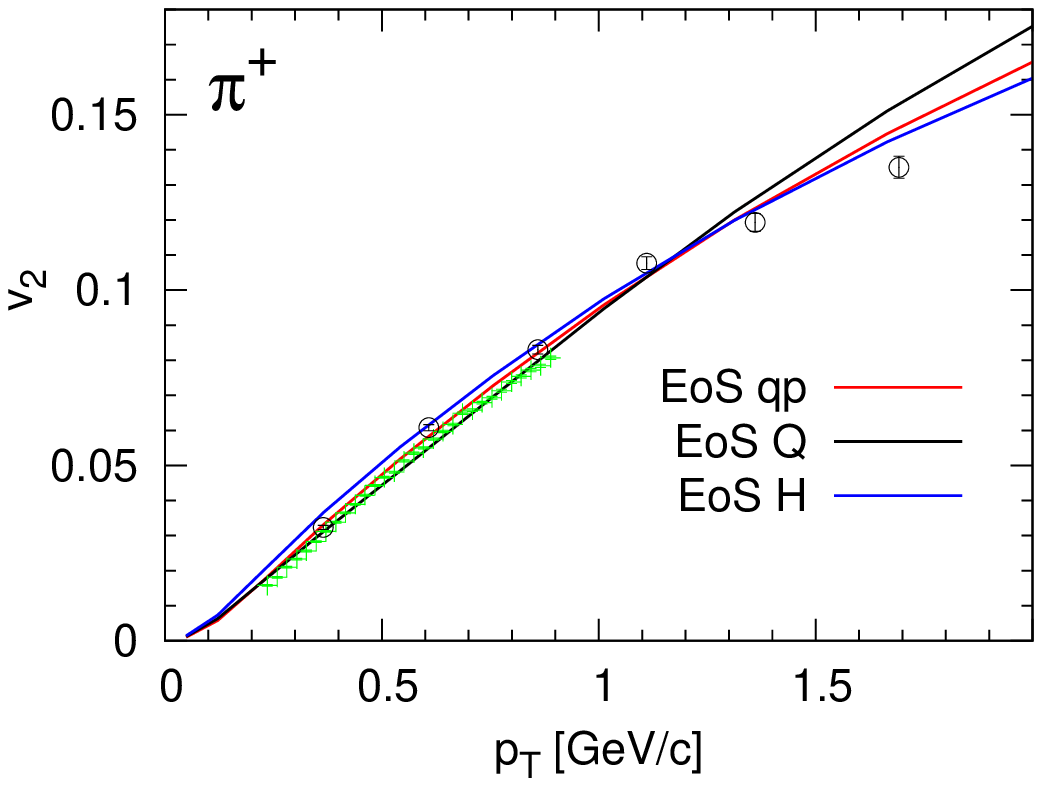}\hspace*{-1mm}
   \includegraphics[width=0.49\textwidth]{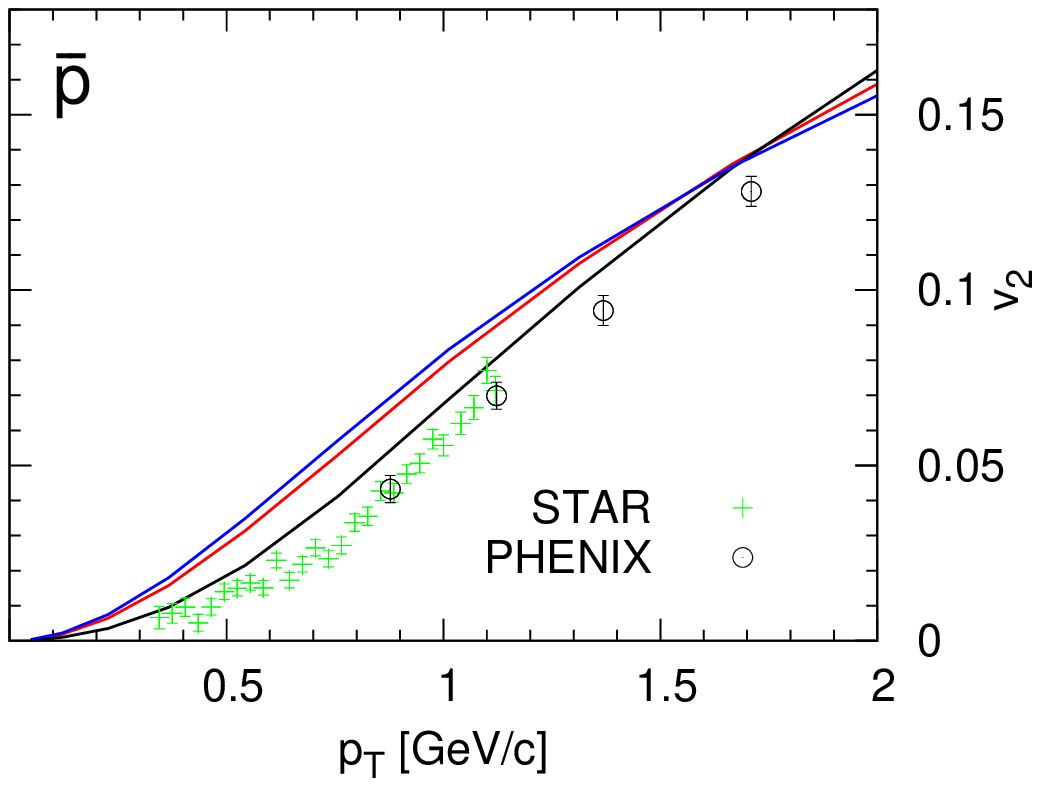}
   \caption{Elliptic flow of pions and antiprotons vs.~transverse
     momentum in minimum bias Au+Au collisions at
     $\sqrt{s_{\mathrm{NN}}}=200$ GeV calculated using three different
     EoSs~\cite{Huovinen:2005gy} and compared with the data by the
     STAR and PHENIX collaborations~\cite{Adams:2004bi,Adler:2003kt}.
     The labels stand for a lattice QCD inspired quasiparticle model
     (qp), EoS with a first order phase transition (Q), and pure
     hadron resonance gas with no phase transition (H). Figure is
     from Ref.~\protect\refcite{confinement}.}
 \label{v2eos}
 \end{center}
\end{figure}

One might want to use the elliptic flow to constrain the EoS after the
initial state and freeze-out temperature are fixed to reproduce the
$p_T$ distributions. Unfortunately elliptic flow is only very weakly
sensitive to the details of the EoS~\cite{Huovinen:2005gy}: The only
flow observable affected by the EoS seems to be the $p_T$-differential
anisotropy of heavy particles, \emph{e.g.}, protons. As shown in
Fig.~\ref{v2eos}, the $v_2(p_t)$ of pions is unchanged within the
experimental errors no matter whether one uses an EoS with (EoS~A) or
without phase transition (EoS~H), or an EoS with a first order phase
transition (EoS~A) or with a smooth crossover (EoS~qp). On the other
hand, the proton $v_2(p_t)$ is sensitive to the EoS, but surprisingly
the EoS with the first order phase transition is closest to the data.
Consequently distinguishing between different lattice QCD EoS
parametrizations is very difficult, see
Ref.~\refcite{Huovinen:2009yb}.

\subsection{Softest point of the Equation of State is quite hard}

The two-particle correlations at low relative momentum provide
information about the space-time structure of the source, and thus
constraints to dynamical models generating the source. For bosons
these correlations are called Bose-Einstein correlations, and for
identical particles the method for their interpretation is called
Hanbury Brown--Twiss (HBT) interferometry, or, as in general case,
femtoscopy~\cite{Lisa:2005dd,Lisa:2008gf,Lisa:2011na}. To make the
measured/calculated three-dimensional correlator easier to understand,
it is often expressed in terms of multi-dimensional Gaussian
parametrization, widths of which are called HBT radii. Reproduction of
the HBT radii measured at RHIC was surprisingly difficult, and for
many years it was not possible to describe simultaneously the particle
spectra, their anisotropies, and HBT radii. An inconsistency referred
to as ``HBT puzzle''~\cite{Pratt:2008qv,Pratt:2009hu}. It turned out
that this inconsistency was due to several small effects. Successful
reproduction of the data required that, 1) the transverse collective
expansion begins very early, 2) the EoS is quite hard, 3) dissipative
effects are included, and 4) contribution from resonance decays is
included~\cite{Pratt:2008qv,Pratt:2009hu,Kisiel:2006is,Broniowski:2008vp}.
When these requirements are taken into account, the present
calculations provide an acceptable fit to the
data~\cite{Werner:2010aa,Bozek:2011ua,Karpenko:2012yf}.

The early build up of collective expansion supports the notion of
early thermalization discussed in Sec.~\ref{thermal}, but the HBT
radii provide as ambiguous support as elliptic flow. If sufficient
collective motion is build up during thermalization, HBT radii can be
fitted also if thermalization takes long (so-called pre-equilibrium
flow)~\cite{Pratt:2008qv,Pratt:2009hu,Broniowski:2008vp}. The
sufficient hardness of the EoS is fortunately more solid
requirement. In practice it means that a bag model EoS with a mixed
phase, where the speed of sound is negligible, is disfavored, and thus
the softest point of the EoS can hardly be any softer than that of
hadron resonance gas~\cite{Chojnacki:2007jc}. A notion in line with
the lattice QCD EoS calculations~\cite{Borsanyi:2010cj}.

Nevertheless, since it is so difficult to constrain the EoS, in the
present calculations the lattice QCD
EoS~\cite{Borsanyi:2010cj,Philipsen:2012nu,Petreczky:2012rq,Borsanyi:2013bia}
is taken as given, and the main interest is in studying the
dissipative properties of the system. At low temperatures the lattice
EoS is equivalent to the EoS of noninteracting hadron resonance gas
(HRG) with all the hadrons and resonances in the Particle Data Book up
to $\sim 2$--2.5 GeV mass. In calculations one often uses an EoS which
is a HRG EoS at low temperature, and connected smoothly to a
parametrized lattice EoS at high
temperature~\cite{Huovinen:2009yb,Bluhm:2013yga}.

In the long run a systematic study of collisions at different energies
may reveal some sensitivity to EoS and thus allow to test
experimentally the lattice QCD prediction. However, since the
parameter space and the amount of data to be fitted are vast, such a
study requires the use of model emulators to map the parameter space
instead of using actual hydrodynamical model to calculate results at
all parameter combinations~\cite{Petersen:2010zt,Novak:2013bqa}.

\subsection{Shear viscosity over entropy density ratio has very low
  minimum}
  \label{etapers}

Once it became clear that ideal fluid dynamics can describe the
particle spectra and their anisotropies quite well, it was
reasonable to assume that the shear viscosity coefficient over entropy
density ratio $\eta/s$ of the matter produced in the collision was
very low. But how low in particular? To answer that required the
development and use of relativistic dissipative hydrodynamics. Of the
other dissipative quantities heat conductivity can be ignored since at
midrapidity the matter formed in the collisions at RHIC and LHC is
almost baryon free. Thus there is no gradient of chemical potential
and no force causing heat flow. The bulk viscosity, on the other hand,
is expected to peak around the phase transition, but be small above
it. If it is small at lower temperatures too, the effect of bulk
viscosity has been evaluated to be smaller than the effect of shear
viscosity~\cite{Song:2009rh}. Large viscosity in hadronic phase can
have a sizable
effect~\cite{Monnai:2009ad,Dusling:2011fd,Noronha-Hostler:2013gga} but
since bulk viscosity in hadron gas is not well known, and there is no
reliable method to distinguish the effects of bulk from the effects of
shear, the former is largely ignored, and the calculations concentrate
on studying the effects of shear viscosity and on extracting the
$\eta/s$ ratio from the experimental data~\cite{Heinz:2013th}.

\begin{figure}
   \includegraphics[width=0.49\textwidth]{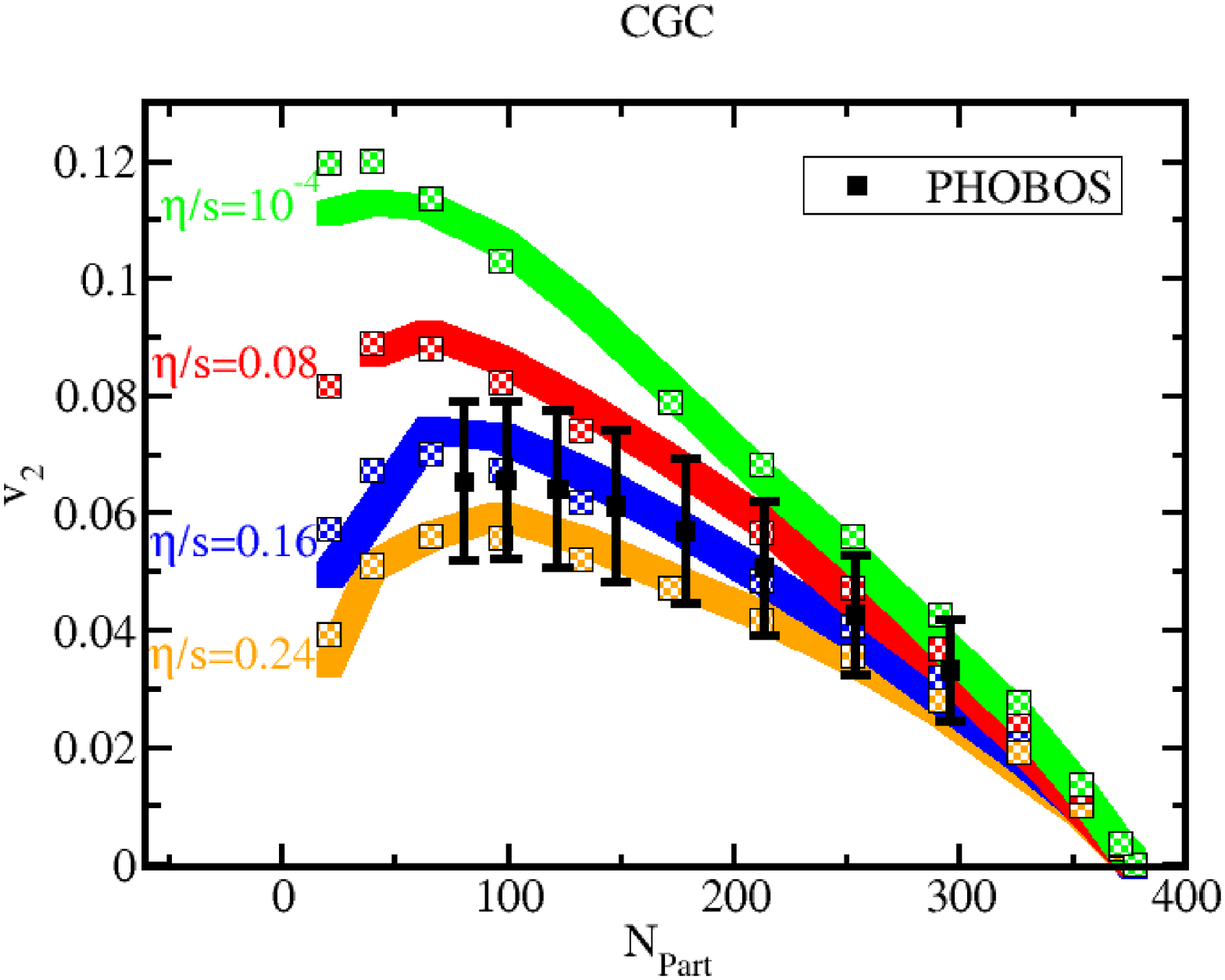}
    \hfill
   \includegraphics[width=0.49\textwidth]{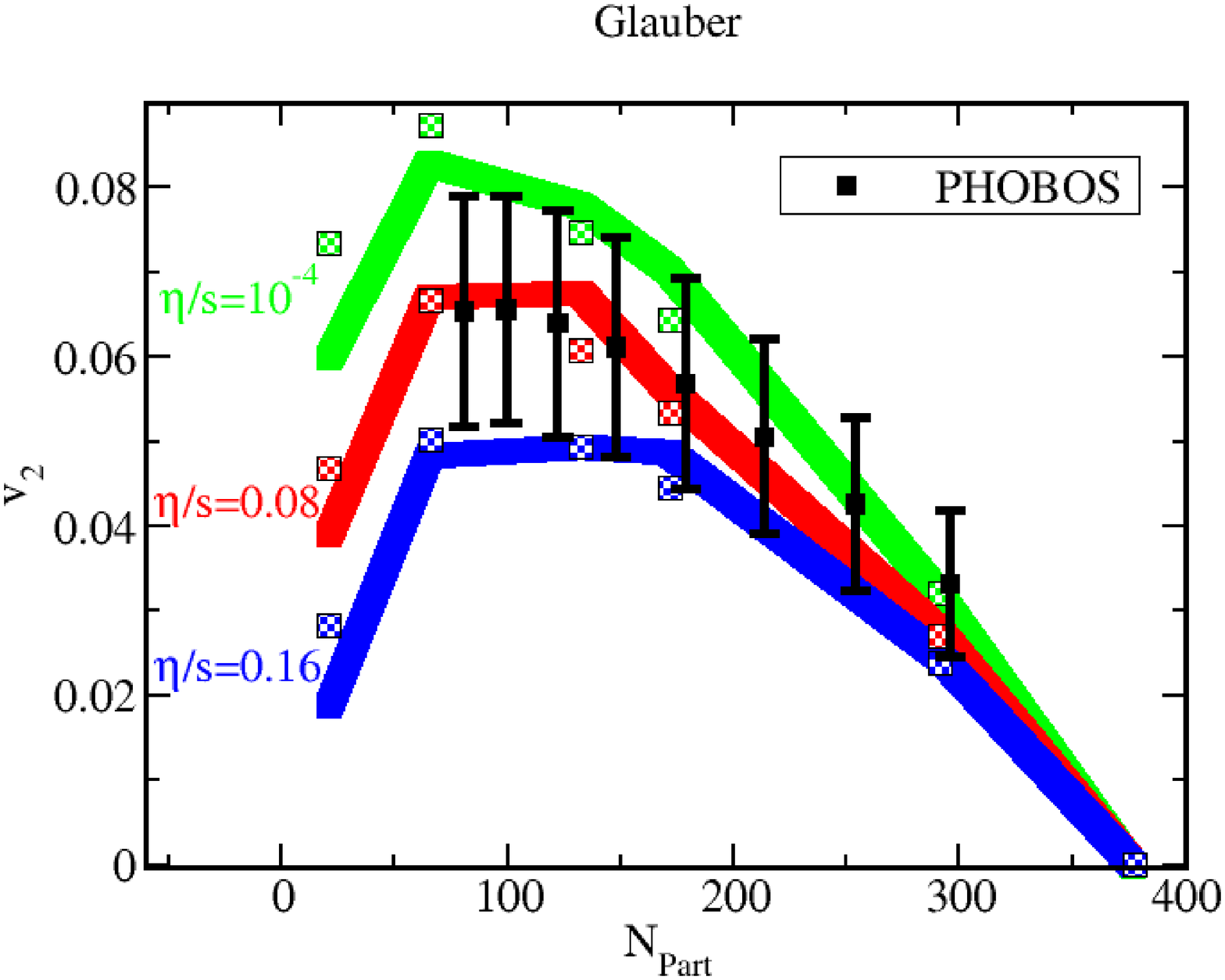}
   \caption{Charged hadron $v_2$ as function of centrality in Au+Au
     collisions at $\sqrt{s_\mathrm{NN}}=200$ GeV using different
     values of $\eta/s$ and KLN (labeled CGC, left) or Glauber (right)
     initial conditions. The larger the number of participants,
     $N_\mathrm{part}$, the more central the collision. The data are
     from Ref.~\protect\refcite{Alver:2007zz}. Reprinted figures with
     permission from Ref.~\protect\refcite{Luzum:2008cw}. Copyright
     (2008) by the American Physical Society.}
 \label{Paul}
\end{figure}

It has been shown that the shear viscosity strongly reduces
$v_2$~\cite{Teaney:2003kp}. Thus in principle extracting the $\eta/s$
ratio from the data is easy: One needs to calculate the $p_T$-averaged
$v_2$ of charged hadrons using various values of $\eta/s$ and choose
the value of $\eta/s$ which reproduces the data. Unfortunately this
approach is hampered by our ignorance of the initial state of the
evolution. Fig.~\ref{Paul} shows a viscous fluid calculation of $v_2$
of charged hadrons from the first attempt to extract $\eta/s$ from the
data. As seen, a curve corresponding to a finite value of $\eta/s$
fits the data best, but the preferred value depends on how the initial
state of hydrodynamic evolution is chosen: Whether one uses
Glauber~\cite{Miller:2007ri} or KLN
approach~\cite{Kharzeev:2001yq,Drescher:2006ca} (see
sect.~\ref{ideal}) causes a factor two difference in the preferred
value ($\eta/s = 0.08$--0.16). Furthermore, the approximations in the
description of the late hadron gas stage in these calculations caused
additional uncertainties, so it was estimated~\cite{Song:2012tv} that
based on these results $\eta/s < 5/(4\pi)$.

The description of hadronic stage in Ref.~\refcite{Luzum:2008cw} was
problematic for two reasons. It assumed chemical equilibrium until the
very end of the evolution, but at RHIC, the final particle yields
correspond to a chemically equilibrated source in $T=160$--165 MeV
temperature~\cite{Andronic:2008gu,Andronic:2009qf}. The decoupling
temperature required to fit the $p_T$-distributions is usually around
$T_\mathrm{dec}=100$--140 MeV\footnote{$T_\mathrm{dec}=140$ MeV in
  Ref.~\refcite{Luzum:2008cw}.}. Thus there must be a stage where the
fluid is in local kinetic, but not in chemical equilibrium. Such a
``chemically frozen'' stage turned out to have surprisingly large
effect on the collective flow in general and elliptic flow in
particular~\cite{Hirano:2002ds,Huovinen:2007xh}. The second problem
was that $\eta/s$ was constant during the entire evolution, but the
theoretical expectation is that hadron gas has much larger viscosity
than what the obtained value $\eta/s = 0.08$--0.16
indicates~\cite{Prakash:1993bt,Demir:2008tr}.

\begin{figure}
 \includegraphics[width=0.9\textwidth]{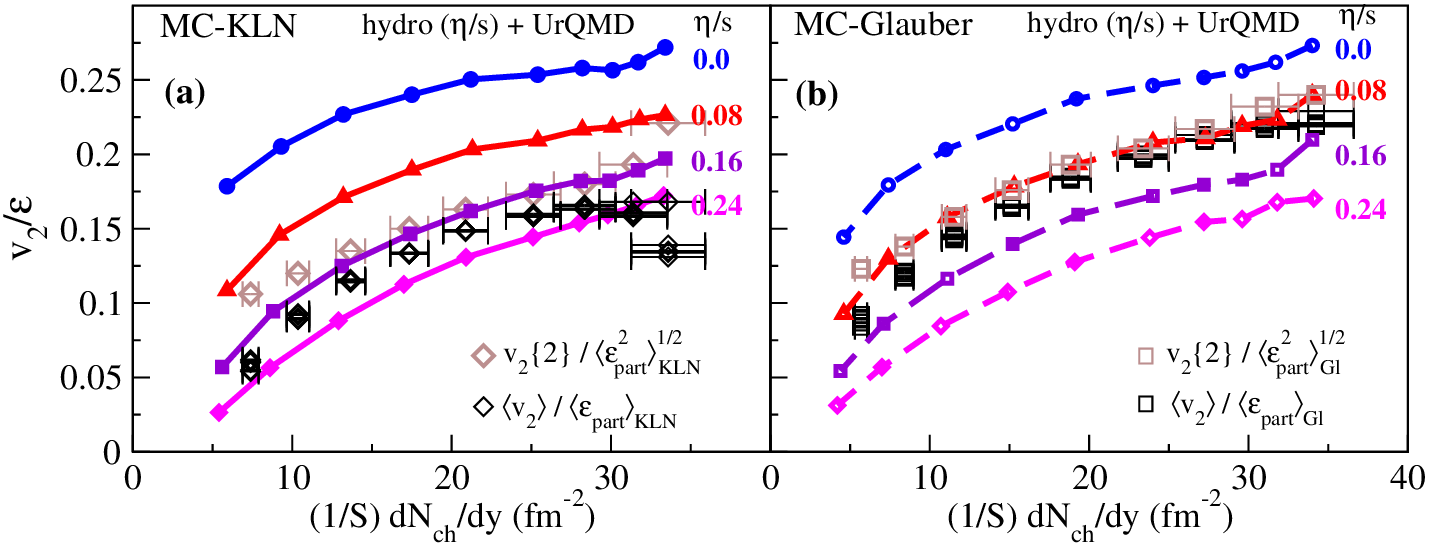}
 \caption{$v_2$ scaled by the initial state anisotropy $\epsilon$,
   $v_2/\epsilon$, as function of centrality, characterized by the
   final multiplicity per area of the initial state, $(1/S)\,\dif
   N/\dif y$, in Au+Au collisions at $\sqrt{s_\mathrm{NN}}=200$
   GeV. The experimental data are from
   Refs.~\protect\refcite{Ollitrault:2009ie},
   \protect\refcite{Adams:2004bi} and~\protect\refcite{Abelev:2008ab}
   for $\langle v_2\rangle$, $v_2\{2\}$, and $\dif N_\mathrm{ch}/\dif
   y$, respectively. The experimental data used in (a) and (b) are
   identical, but the normalization factors
   $\langle\varepsilon_\mathrm{part}\rangle$,
   $\langle\varepsilon_\mathrm{part}^2 \rangle^{1/2}$ and $S$ are
   taken from the MC-KLN model in (a) and from the MC-Glauber model in
   (b). The viscous hydro + UrQMD hadron cascade hybrid model
   calculations from Ref.~\protect\refcite{Song:2010mg} are done using
   the MC-KLN initial conditions in (a) and MC-Glauber in (b). 
   Reprinted figure with permission form
   Ref.~\protect\refcite{Song:2010mg}. Copyright (2011) by the
   American Physical Society.}
 \label{huichao}
\end{figure}

In subsequent calculations these uncertainties have been addressed. A
state-of-the-art calculation of Ref.~\refcite{Song:2010mg} employs a
hybrid viscous hydro + UrQMD hadron cascade model. Such a model
reduces the uncertainty related to the description of hadron gas since
it provides a dynamical description of chemistry and dissipative
properties---including bulk viscosity---based on the scattering cross
sections of various hadron species. Unfortunately some uncertainties
still remain: It is uncertain whether hadron cascade is applicable in
the vicinity of phase transition where the switch from hydro to
cascade is done, the dissipative properties of the system (\emph{i.e.}
$\eta$, $\zeta$, $\tau_\pi$ and $\tau_\Pi$) change abruptly at the
switch, and the results also depend on when the switch from hydro to
cascade is done~\cite{Song:2010aq}. The elliptic flow coefficient
$v_2$ as function of centrality as calculated in
Ref.~\refcite{Song:2010mg} is shown in Fig.~\ref{huichao}.  In this
figure the coefficients $v_2$ have been scaled by the anisotropy of
the initial shape, $\epsilon$, and consequently the resulting
$v_2/\epsilon$ is almost independent of the
initialization. Unfortunately $\epsilon$ is not a measurable, but a
model dependent quantity. Thus the data has to be scaled by the same
$\epsilon$ which was used in the calculation, and the data points in
Fig.~\ref{huichao} depend on the model used to initialize the
calculation. The result is almost identical to the one shown in
Fig.~\ref{Paul}---Glauber initialization favors lower value of
viscosity, $\eta/s \approx 0.08$, than KLN initialization, $\eta/s
\approx 0.16$. Since the uncertainties are smaller, this result was
estimated to provide a limit $1/(4\pi) < \eta/s < 2.5/(4\pi)$ for the
effective QGP viscosity~\cite{Song:2012tv,Song:2010mg}. For further
discussion of evaluating $\eta/s$, see Ref.~\refcite{Song:2012tv}.

\section{What we are working with}
 \label{now}

 \subsection{$\eta/s(T)$}

\begin{figure}
    \hfill
   \includegraphics[width=0.45\textwidth]{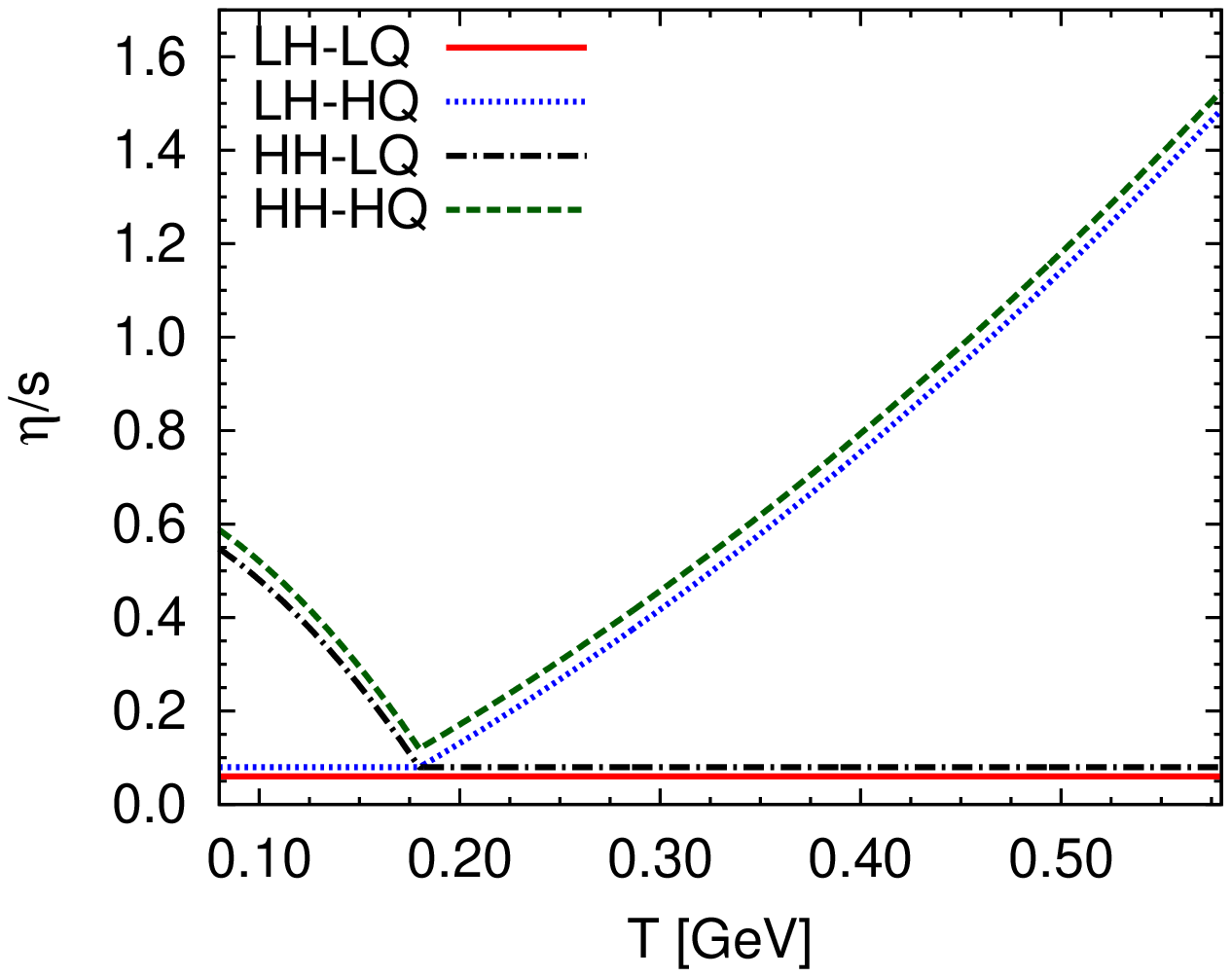}
    \hfill
   \includegraphics[width=0.45\textwidth]{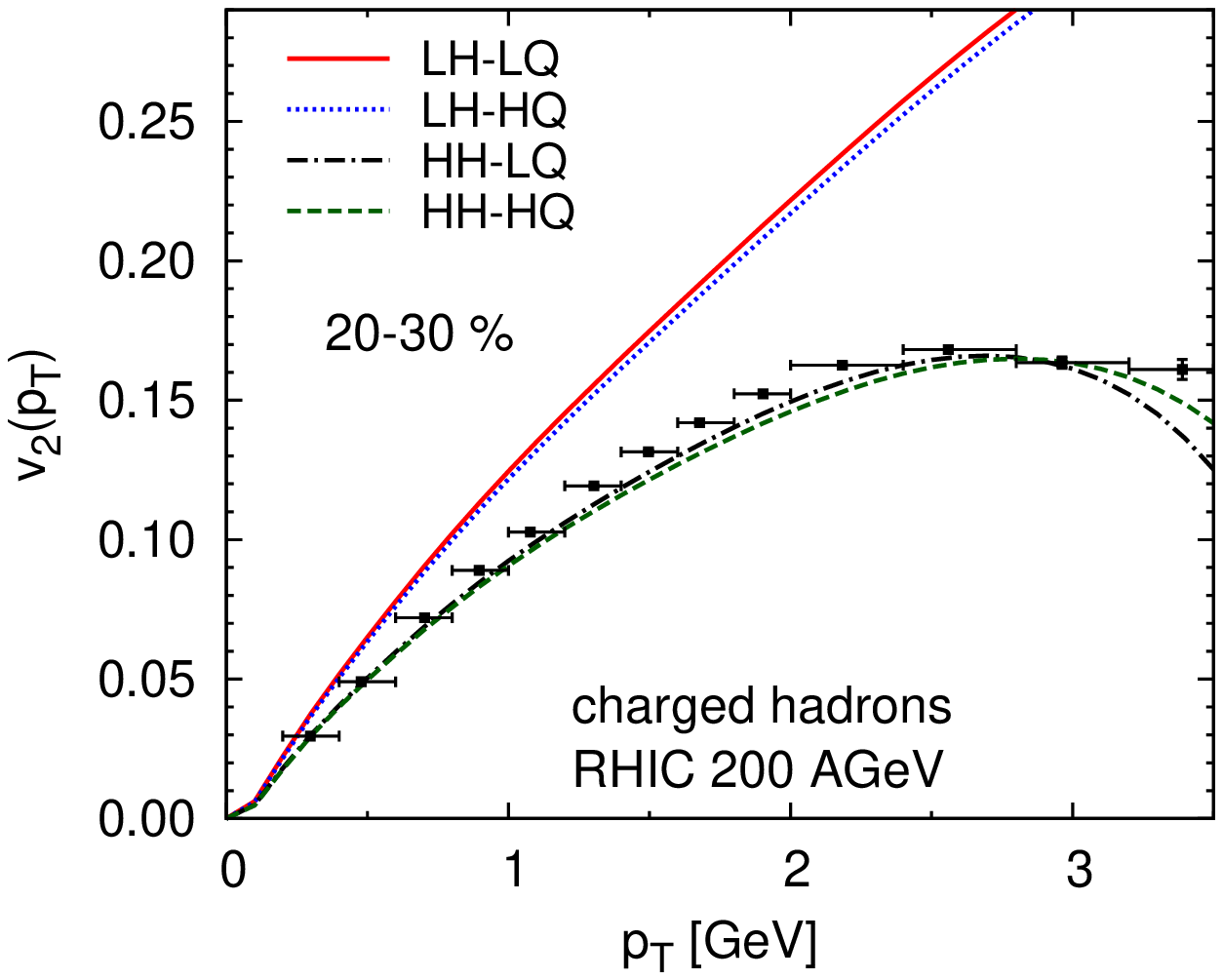}
    \hspace*{5mm}
\\

    \hfill
   \includegraphics[width=0.45\textwidth]{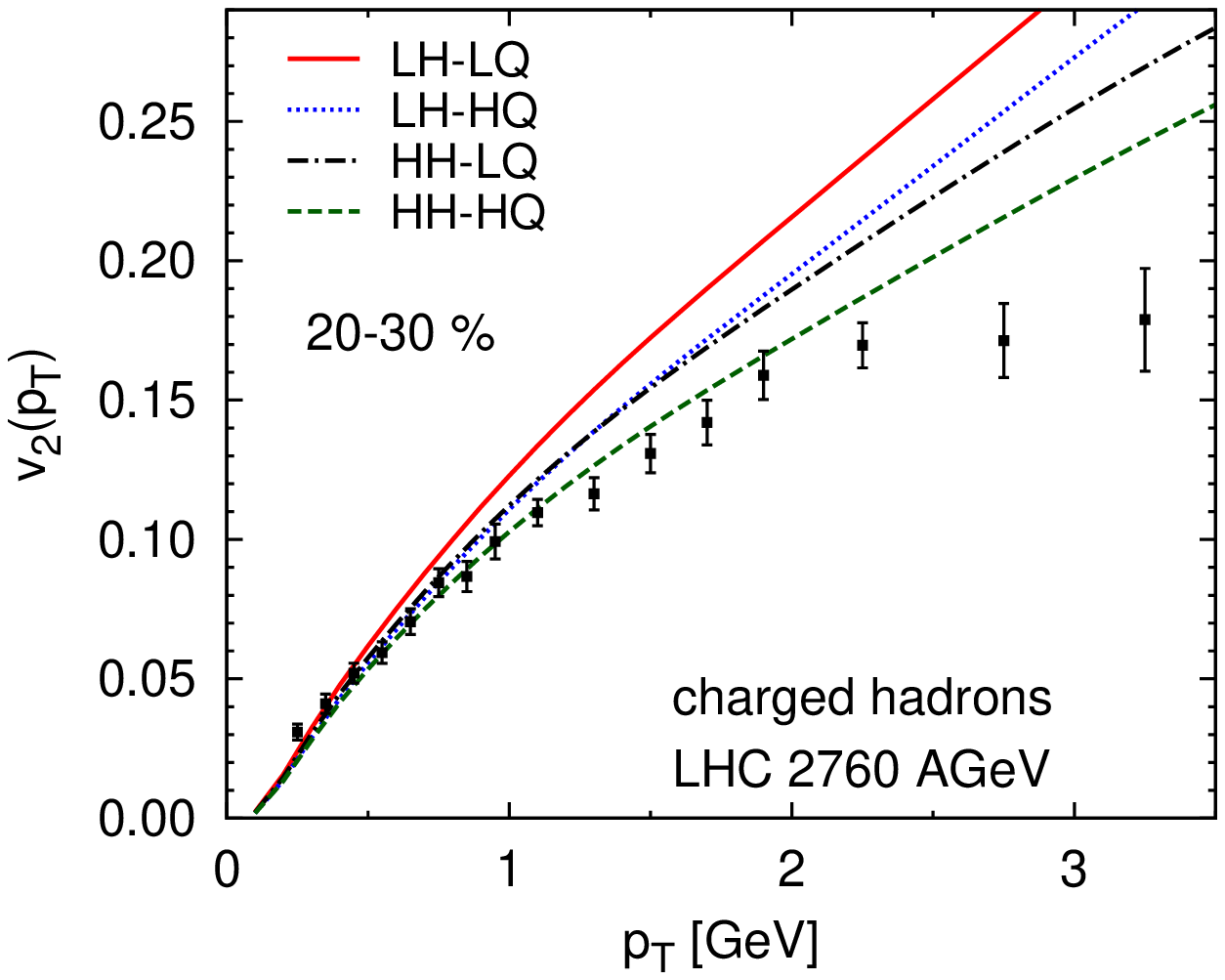}
    \hfill
   \includegraphics[width=0.45\textwidth]{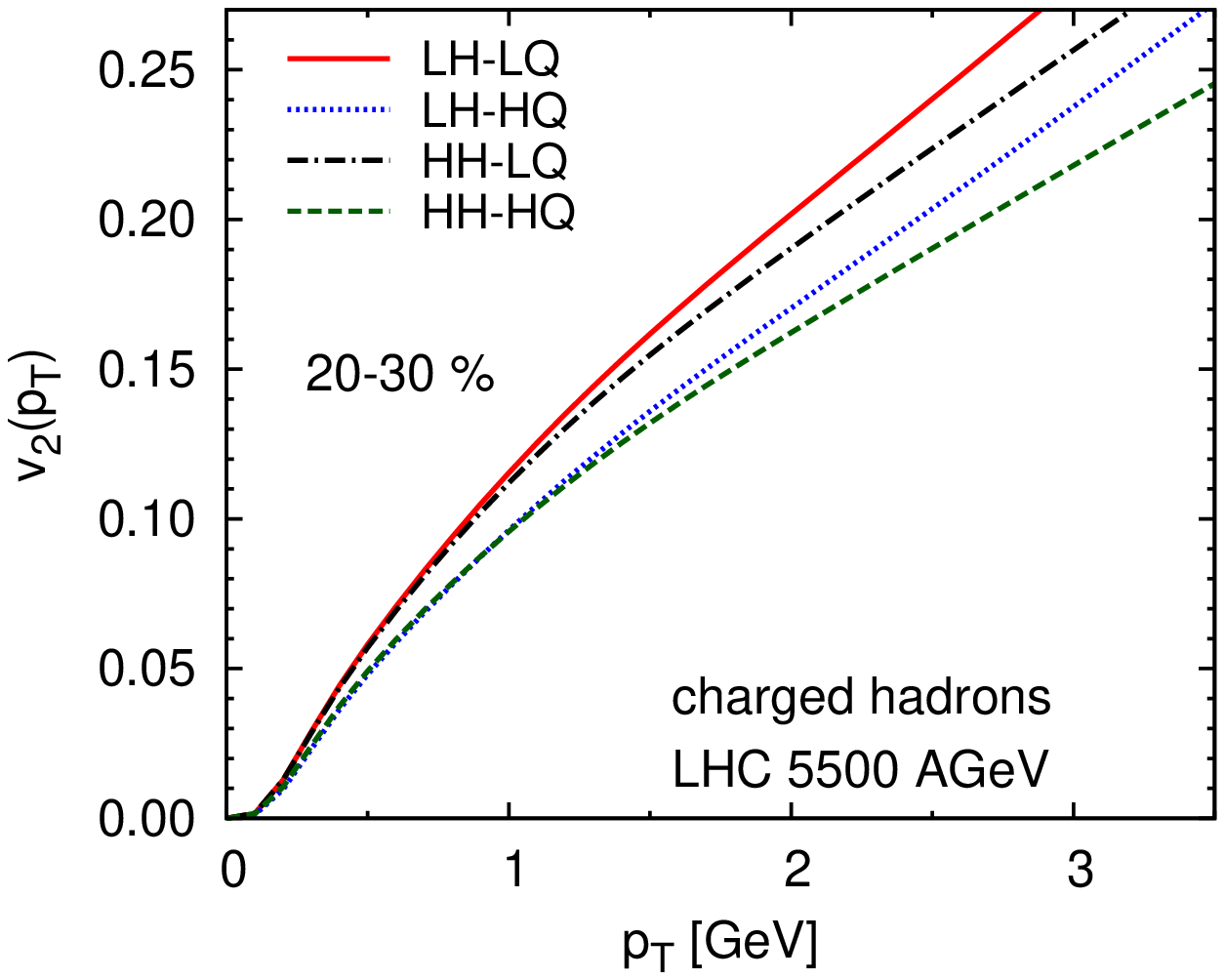}
    \hspace*{5mm}
    \caption{(top left:) Different parametrizations of $\eta/s$ as a
      function of temperature.  The (LH-LQ) line is shifted downwards
      and the (HH-HQ) line upwards for clarity. Labels refer to low
      (L) or high (H) viscosity in the hadronic (H) or partonic (Q)
      phases. (top right:) $v_2(p_T)$ of charged hadrons in the
      20-30\% Au+Au collisions at $\sqrt{s_\mathrm{NN}} = 200$ GeV
      (RHIC). Data are from Refs~\protect\refcite{Bai,Tang:2008if}.
      (bottom left:) $v_2(p_T)$ of charged hadrons in the 20-30\%
      Pb+Pb collisions at $\sqrt{s_\mathrm{NN}} = 2.76$ TeV
      (LHC). Data are from Ref~\protect\refcite{Lissu}. (bottom
      right:) $v_2(p_T)$ of charged hadrons in the 20-30\% Pb+Pb
      collisions at $\sqrt{s_\mathrm{NN}} = 5.5$ TeV (LHC). All the
      figures are from Ref.~\protect\refcite{sQM}.}
 \label{etas1}
\end{figure}

Calculations discussed in section~\ref{etapers} assumed that the
$\eta/s$-ratio is constant. We know no fluid where $\eta/s$ would be
temperature independent, and there are theoretical reasons to expect
it to depend on temperature with a minimum around the phase
transition~\cite{Joe}. Thus the temperature independent $\eta/s$ is an
effective viscosity, and its connection to the physical, temperature
dependent, shear viscosity coefficient is unclear. What complicates
the determination of the physical shear viscosity coefficient, is that
the sensitivity of the anisotropies to dissipation varies during the
evolution of the system. As studied in Ref.~\refcite{Harri}, and
illustrated in Fig.~\ref{etas1}, at RHIC ($\sqrt{s_\mathrm{NN}}=200$
GeV) $v_2(p_T)$ is insensitive to the value of $\eta/s$ above phase
transition, but very sensitive to its value in the hadronic phase. At
the present LHC energy ($\sqrt{s_\mathrm{NN}}=2.76$ TeV) the shear
viscosity in the plasma phase does affect the final $v_2(p_T)$, but
not more than the shear viscosity in the hadronic phase. It is only at
the full LHC energy, $\sqrt{s_\mathrm{NN}}=5.5$ TeV where the
viscosity in the plasma phase dominates, and dissipation in the
hadronic phase has only a minor effect. Note that a change of the
minimum value of $\eta/s$ would clearly change $v_2(p_T)$ at all
energies.

\begin{figure}
    \hfill
   \includegraphics[width=0.46\textwidth]{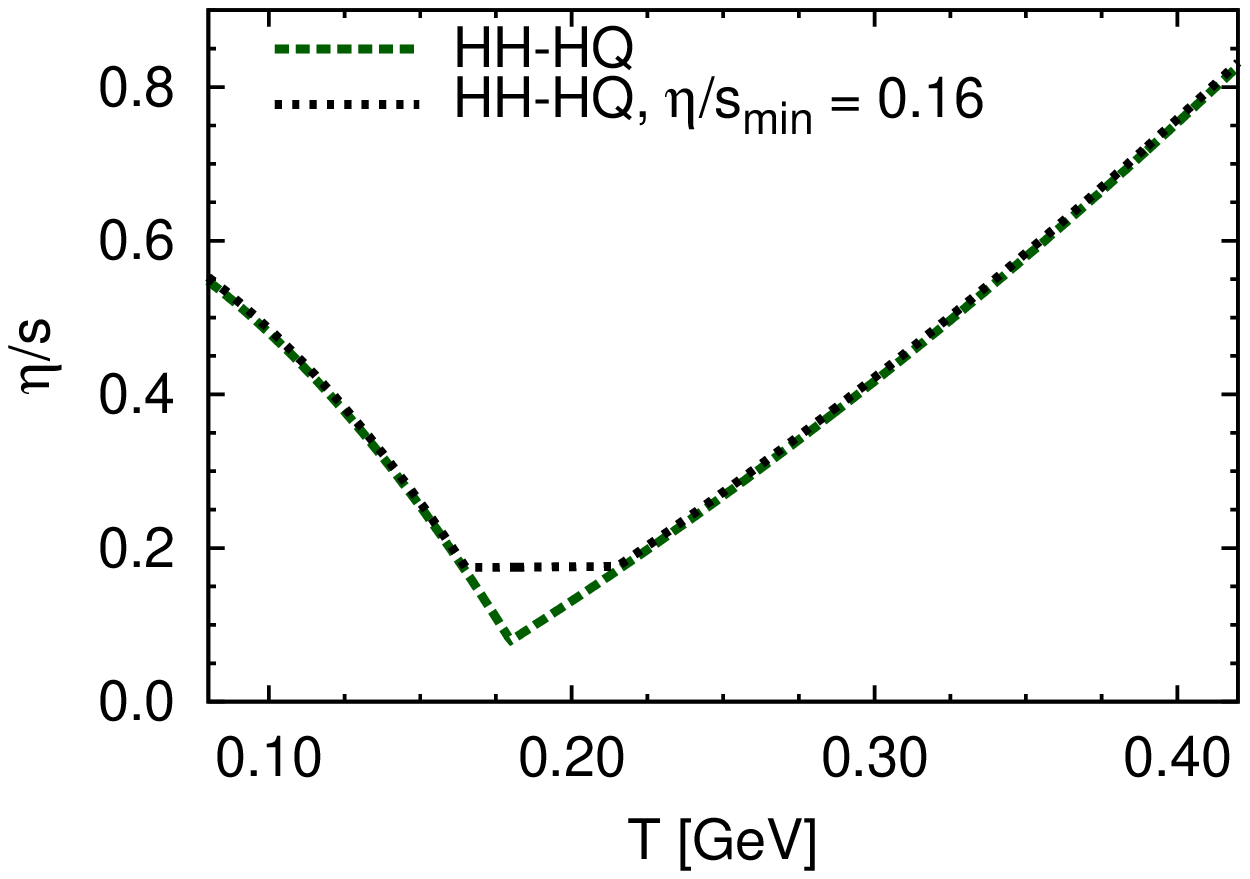}
    \hfill
   \includegraphics[width=0.42\textwidth]{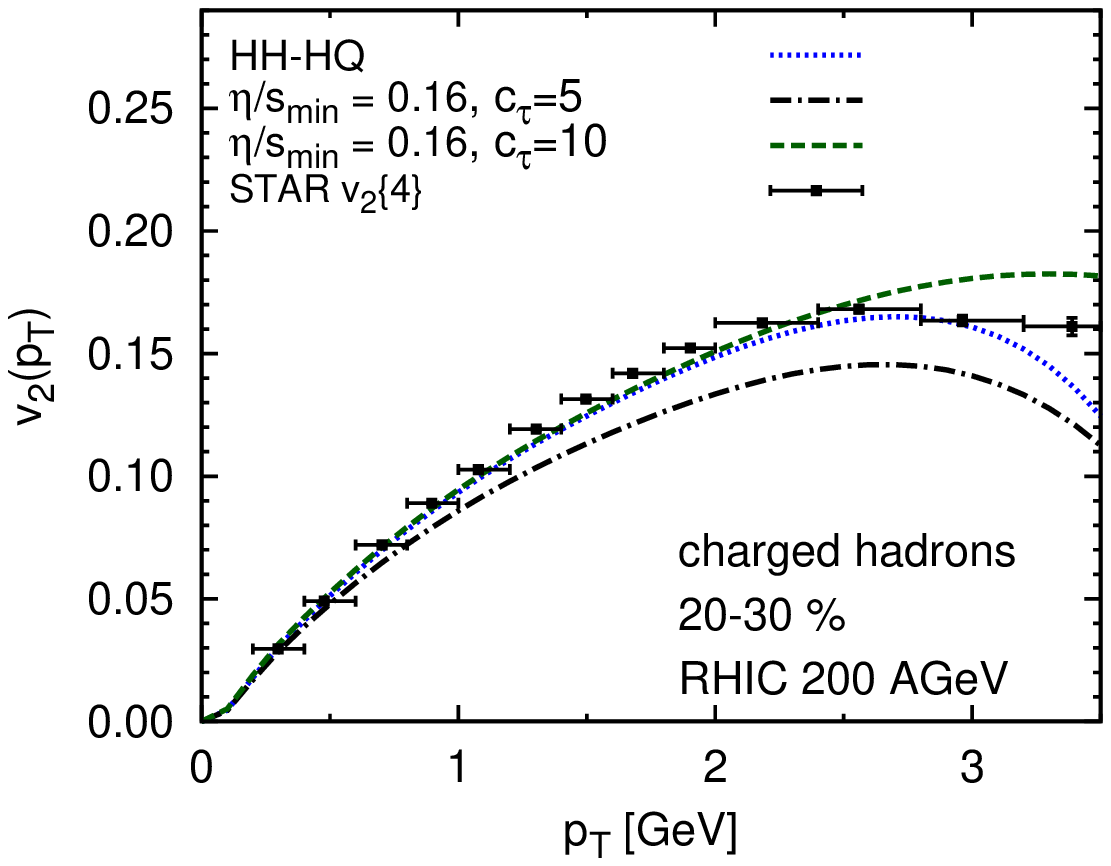}
    \hspace*{5mm}
    \caption{(left:) Parametrizations of $\eta/s$ as a function of
      temperature with different minima. The (HH-HQ) line is the same
      than in Fig.~\protect\ref{etas1}. (right:) $v_2(p_t)$ of charged
      hadrons at RHIC using $\eta/s(T)$ with different minima and
      different relaxation times. Figures are from
      Ref.~\protect\refcite{Harri}.}
 \label{etas2}
\end{figure}

Additional factor complicating the determination of the temperature
dependence of $\eta/s$ is that the effect of viscosity on the
anisotropies does not depend only on the ratio $\eta/s$, but also on
the relaxation time $\tau_\pi$ of the shear stress tensor. This is
demonstrated in Fig.~\ref{etas2}. If the minimum value of $\eta/s$ is
increased by a factor two, $v_2$ is reduced as expected, but if the
relaxation time is also increased by a factor two, the effect of the
increase in $\eta/s$ is almost completely compensated~\cite{Harri}.
The interplay of relaxation time and shear viscosity was discussed
also in Refs.~\refcite{Song:2010aq} and~\refcite{Shen:2011kn}, where it
was found that to reproduce hybrid model results using viscous
hydrodynamics only, it is not sufficient to increase $\eta/s$ in the
hadronic phase, but one should also change the relaxation time.

So far we have seen that calculations with constant $\eta/s$ require
slightly larger value of $\eta/s$ at LHC than at
RHIC~\cite{Song:2011qa,Gale:2012rq}. This is in line with the increase
of $(\eta/s)(T)$ in high temperatures, but as shown in
Ref.~\refcite{Song:2011qa}, one cannot uniquely constrain $(\eta/s)(T)$
by fitting the spectra and $v_2$ alone.

 \subsection{Fluctuations}

\begin{figure}[b]
  \begin{minipage}[t]{0.49\textwidth}
    \includegraphics[width=\textwidth]{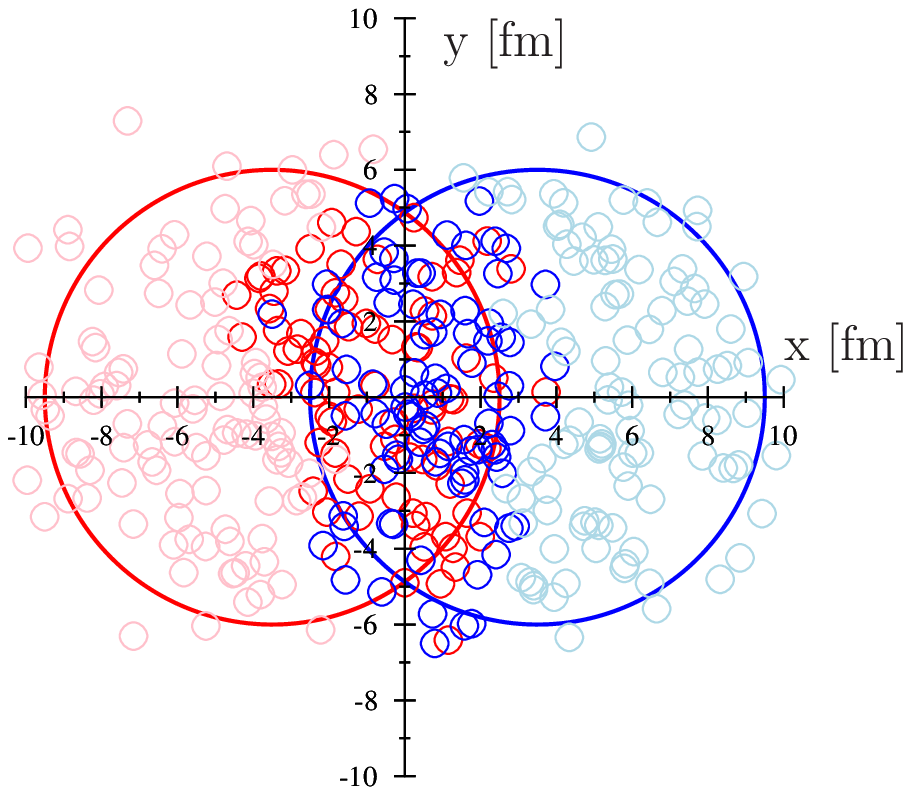}
    \caption{An example of the positions of interacting nuclei in
      MC-Glauber model. Figure is from Ref.~\protect\refcite{Hannu}.}
    \label{fluktu}
  \end{minipage}
   \hfill
  \begin{minipage}[t]{0.49\textwidth}
    \includegraphics[width=\textwidth]{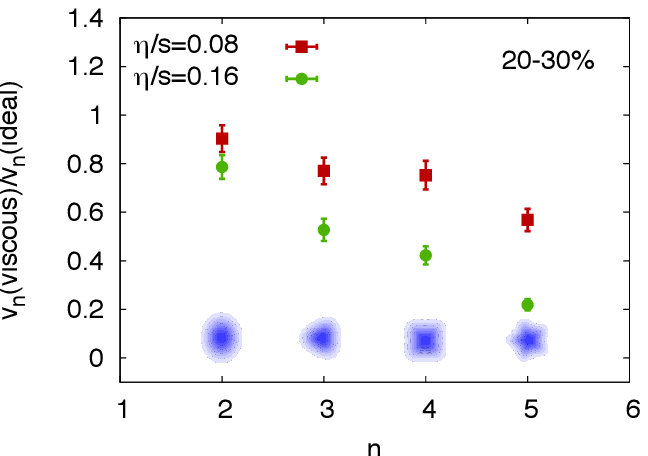}
    \caption{Ratio of the anisotropy coefficients of charged hadrons
      in viscous calculation to the coefficients in ideal fluid
      calculation~\cite{Nalle}. Figure shown in
      Ref.~\cite{confinement} by courtesy of Bj\"orn Schenke.}
    \label{vn}
  \end{minipage}
\end{figure}

In the average the matter formed in heavy-ion collisions has a smooth
shape as indicated in Fig.~\ref{schematic}, but in each event that is
not true. Nuclei are not smooth distributions of nuclear matter, but
consist of individual nucleons, and thus the collision zone can be
expected to have a highly irregular shape which fluctuates
event-by-event, see Fig.~\ref{fluktu}. The initial state models (see
sect.~\ref{ideal}) can be generalized to so-called Monte-Carlo (MC)
models to take this into account. In those models the initial
Woods-Saxon distributions of nuclear matter are sampled to generate
configurations of nucleons in nuclei, and the number of
participants/collisions~\cite{Miller:2007ri} or gluon
production~\cite{Drescher:2006ca,Drescher:2007ax} in an event is
calculated based on the positions of individual nuclei.

It was shown already a while ago that the final observables differ
whether one first averages the initial state, and evolves it
hydrodynamically, or one evolves the event-by-event fluctuating
initial states individually, and averages the
results~\cite{Aguiar:2001ac,Socolowski:2004hw}. This became widely
recognized only years later when it was realized that because of the
irregular shape of each event, not only even, but also odd anisotropy
coefficients $v_n$ are finite and measurable~\cite{AlverRoland}. The
higher coefficients are very helpful for extracting the transport
coefficients from the data, since the larger the $n$, the more
sensitive the coefficient $v_n$ is to viscosity~\cite{Nalle}, see
Fig.~\ref{vn}. Thus the study of fluctuations provides a way to
distinguish different initializations, and first results for the
$p_T$-dependence of $v_2$ and $v_3$ (called triangular flow) seem to
favor the MC-Glauber initialization~\cite{Qiu:2011hf}.

It has been suggested that initial fluctuations provide also a way to
circumvent our ignorance of the initial shape~\cite{Luzum:2012wu}. In
most central collisions the anisotropies are entirely driven by
fluctuations, which are better understood than the average shape in
semi-central collisions. Thus evaluating $\eta/s$ using $v_n$ in the
most central collisions should be less sensitive to the model used to
calculate the initial state. The preliminary result of such an
analysis was that $0.07 < \eta/s < 0.43$~\cite{Luzum:2012wu}. This
value is again for effective viscosity which does not exclude the
possibility that $\eta/s$ could be even smaller or larger in some
temperature region. For a further discussion of flow and viscosity,
see Ref.~\refcite{Heinz:2013th}.

\begin{figure}
 \begin{center}
    \hfill
   \includegraphics[width=0.44\textwidth]{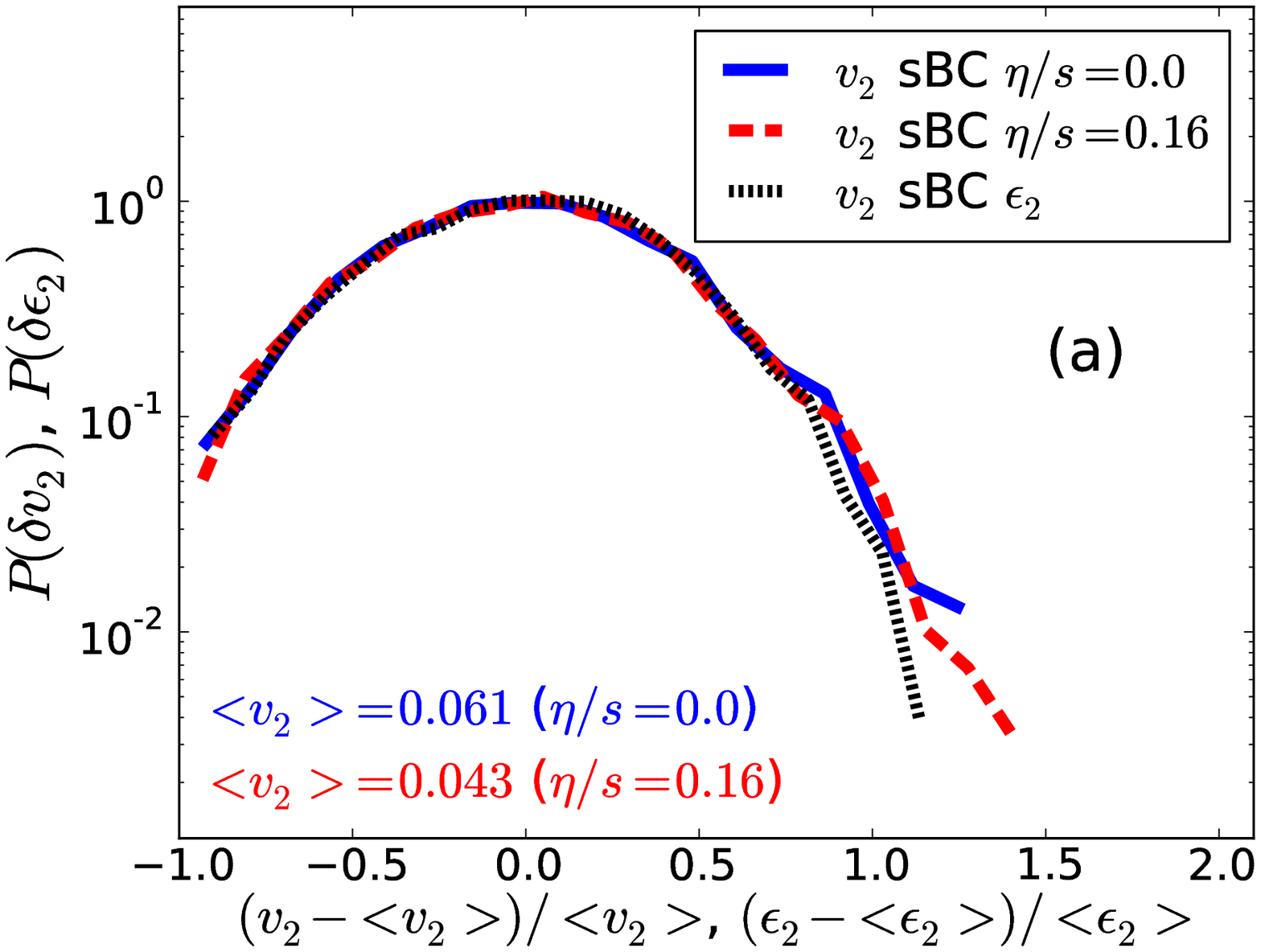}
    \hfill
   \includegraphics[width=0.44\textwidth]{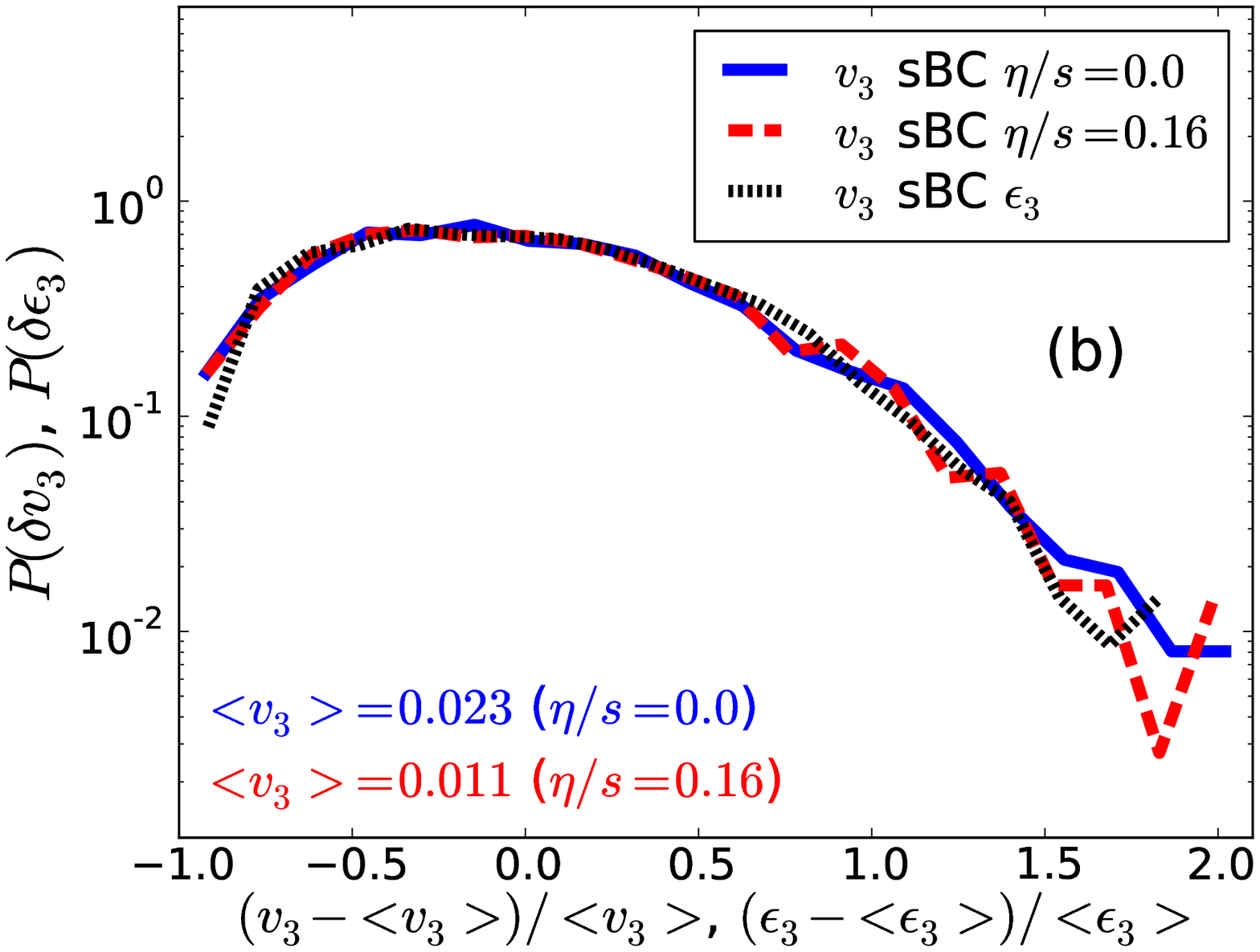}
    \hspace*{4mm}
   \caption{Probability distributions: a) P($\delta v_2$) and
     P($\delta \epsilon_2$), and b) P($\delta v_3$) and
     P($\delta\epsilon_3$ ) in the 20 − 30 \% centrality class with
     sBC Glauber model initialization and two different values of
     $\eta/s$, $\eta/s = 0$ and $\eta/s = 0.16$.  
     $\delta v_n = (v_n-\langle v_n \rangle)/\langle v_n \rangle$ and
     $\delta \epsilon_n = 
     (\epsilon_n-\langle\epsilon_n\rangle)/\langle\epsilon_n\rangle$.
     Figures are from Ref.~\cite{Niemi:2012aj}.}
 \label{deltavee}
 \end{center}
\end{figure}

In event-by-event studies it is not sufficient to reproduce only the
average values of $v_n$, but the fluctuations of the flow coefficients
should be reproduced as well. The distributions of these fluctuations
provide a way to constrain the fluctuation spectrum of initial state
models independently of the dissipative properties of the fluid. As
shown in Fig.~\ref{deltavee}, once the average $v_n$ has been scaled
out, the distributions of these fluctuations, \emph{i.e.},
$(v_n-\langle v_n \rangle)/\langle v_n \rangle$ or 
$v_n/\langle v_n \rangle$, are almost independent of viscosity. The
independence extends to other details of the evolution to such an
extent, that the distributions of the fluctuations of initial
anisotropies are good approximations of the measured distributions of
$v_n$~\cite{Niemi:2012aj}, and thus it is sufficient to compare the
fluctuations of initial shape, $\epsilon_n$, to the observed
fluctuations of $v_n$. Unlike the recently developed IP-Glasma
model~\cite{ipglasma,ipglasma2}, neither MC-Glauber nor MC-KLN model
seems to be able to reproduce the measured
fluctuations~\cite{Jia:2012ve}.

\subsection{Initialization}

During last two years there have been major advances in modeling the
initial state of hydrodynamical evolution. So called IP-Glasma
model~\cite{ipglasma,ipglasma2} is based on Color Glass Condensate and
employs the IP-Sat (Impact Parameter dependent Saturation)
model~\cite{Bartels:2002cj,Kowalski:2003hm} of nucleon wavefunctions
to generate fluctuating gluon fields in the initial collision, and the
classical Yang-Mills dynamics to evolve these
fields~\cite{Kovner,Kovchegov:1997ke,Krasnitz:1,Krasnitz:2,Krasnitz:3,Lappi}.
Unlike most of the initial state models, the IP-Glasma model includes
the fluctuations of color charge in colliding nucleons too, not only
the fluctuations of nucleon positions. Another advantage is that the
model includes some of the pre-equilibrium dynamics of the gluon
fields making the model less sensitive on the time when one switches
to hydrodynamics~\cite{Gale:2012rq}. Unfortunately the description is
still incomplete: the present IP-Glasma model does not lead to a
thermal system, but final thermalization still has to be
assumed. Nevertheless, calculations using the IP-Glasma initialization
reproduce both the fluctuations and the average values of $v_2$, $v_3$
and $v_4$~\cite{Gale:2012rq,Gale:2012in}, which makes this approach
very promising. For a general overview of IP-Glasma, fluctuations and
hydrodynamics, see Ref.~\refcite{Gale:2013da}.

\section{What worries us}

As described in the previous sections, fluid dynamics has been very
successful in describing heavy ion collisions. However, at the time of
this writing there are some data which may cause difficulties for the
conventional fluid dynamical picture.

 \subsection{Photons}

\begin{figure}
 \begin{minipage}[t]{0.48\textwidth}
  \includegraphics[width=\textwidth]{v2_alice.eps}
  \caption{Thermal photon $v_2$ vs.~$p_T$ for 0--40\% central
    collisions at LHC. The preliminary direct photon data are from
    Ref.~\protect\refcite{Lohner:2012ct}.  Reprinted figure with
    permission from
    Ref.~\protect\refcite{Chatterjee:2013naa}. Copyright (2013) by the
    American Physical Society.}
  \label{photonproblem}
 \end{minipage}
  \hfill
 \begin{minipage}[t]{0.48\textwidth}
  \includegraphics[width=0.95\textwidth]{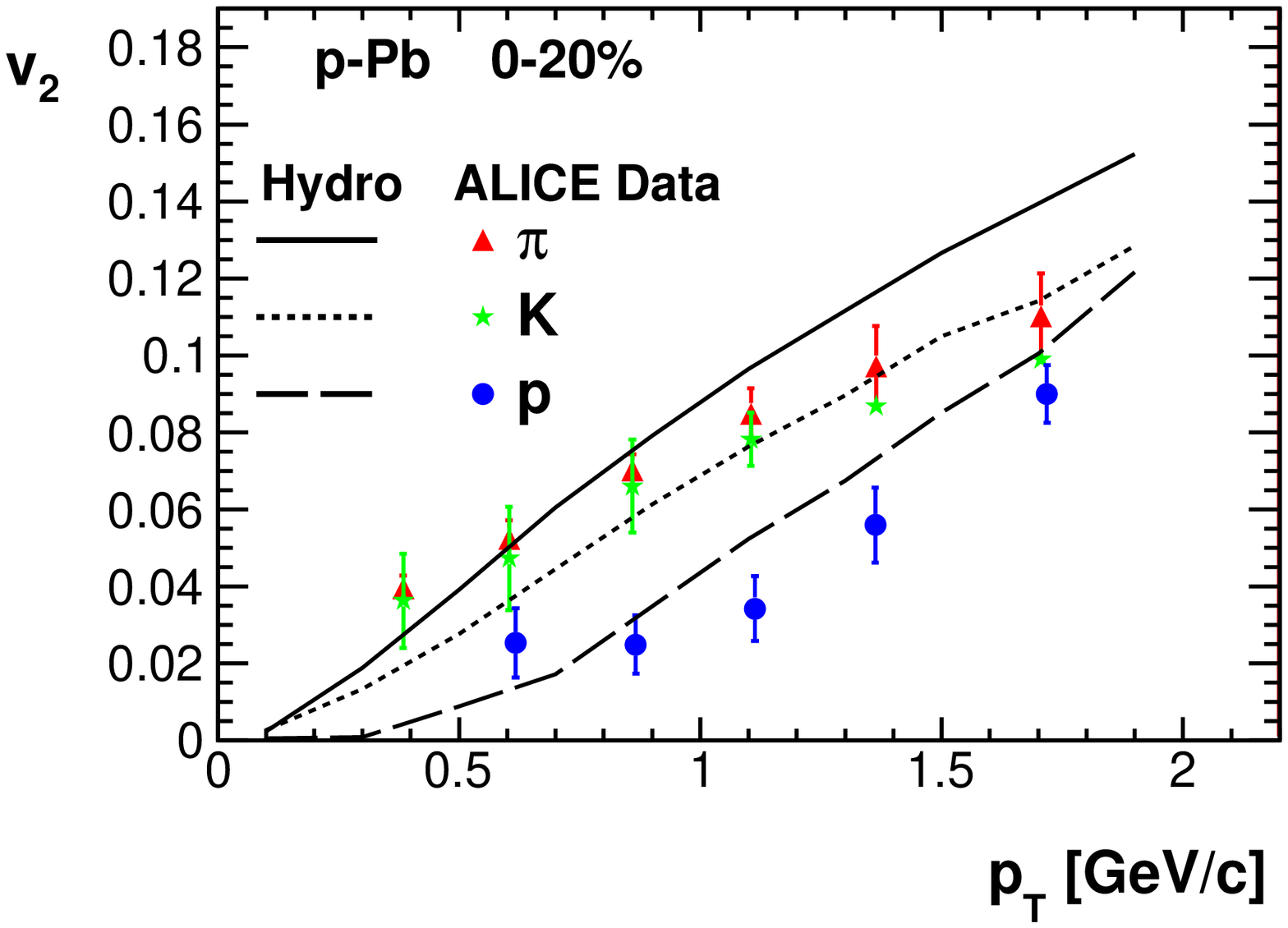}
  \caption{$v_2$ vs.~$p_T$ for identified particles in 0--20\% most
    central $p+Pb$ collisions. Data are from
    Ref.~\protect\refcite{ABELEV:2013} and the hydrodynamical
    calculation from Ref.~\protect\refcite{Bozek:2013ska}. Figure
    courtesy by Piotr Bozek.}
  \label{pPb}
 \end{minipage}
\end{figure}

Unlike hadrons and partons, photons and leptons interact only
electromagnetically, and hardly scatter at all after being
produced. Thus the observed photon and lepton spectra contain
contributions from all stages of a heavy-ion collision, and can be
used to probe the early hot and dense stage~\cite{Kajantie:1981wg}.
The yield and $p_T$-spectrum of photons in heavy-ion collisions is
fairly well
understood~\cite{Turbide:2005fk,Turbide:2007mi,Chatterjee:2008tp,Chatterjee:2011dw,Holopainen:2011pd}
---see also Ref.~\refcite{Gale:2012xq} and references therein---but the
recent measurements of direct photon $v_2$ at RHIC~\cite{Adare:2011zr}
and LHC~\cite{Lohner:2012ct} have presented a puzzle: If photons are
emitted during all stages of the evolution, and momentum anisotropy
increases during the evolution, photon $v_2$ should be smaller than
the hadron $v_2$. Such a behavior is seen in the theoretical
calculations~\cite{Chatterjee:2005de}, but not in the data where the
photon $v_2(p_T)$ is roughly equal to the observed pion
$v_2(p_T)$~\cite{Adare:2011zr,Lohner:2012ct}. The further refinements
of the calculations, where event-by-event
fluctuations~\cite{Chatterjee:2013naa} and viscosity have been taken
into account~\cite{Dion:2011pp,Shen:2013cca}, have not been able to
increase the $v_2$ to the observed level, see
Fig.~\ref{photonproblem}. So far the only approach to get close to the
data has used parametrized expansion with very strong initial
expansion and large photon production rates in the hadronic
phase~\cite{vanHees:2011vb}. Thus the reproduction of the data remains
a challenge to our understanding of microscopic production rates
and/or expansion dynamics.

 \subsection{$p+Pb$ collisions}

In $p+Pb$ collisions at $\sqrt{s_\mathrm{NN}} = 5.02$ TeV, the
measured dihadron
correlations~\cite{Abelev:2012cya,Chatrchyan:2013nka,Aad:2013fja,ABELEV:2013},
multiplicity and species dependence of average
$p_T$~\cite{Abelev:2013bla,Chatrchyan:2013eya}, and elliptic and
triangular flows~\cite{Chatrchyan:2013nka,Aad:2013fja,ABELEV:2013} all
depict features easily explained by using hydrodynamics
\cite{Bozek:2012gr,Bozek:2013uha,Pierog:2013,Qin:2013,Werner:2013,Bozek:2013ska}
(For a short summary see Ref.~\refcite{Bozek:2013yfa}). Especially
striking is the mass ordering of the $p_T$-differential elliptic
flow~\cite{ABELEV:2013}, see Fig.~\ref{pPb}. As discussed in
section~\ref{thermal}, this was taken as a strong indicator of the
formation of a thermal system. However, it is questionable whether
hydrodynamics is applicable to such a small system. Gradients are so
large, that dissipative corrections should be of the same order than
equilibrium pressure. As discussed in Ref.~\refcite{Bzdak:2013zma},
large dissipative corrections are a problem even in $Pb+Pb$
collisions, but in $Pb+Pb$ collisions corrections are large only for a
small fraction of the lifetime of the system, whereas corrections are
large for a significant fraction of the lifetime in $p+Pb$ collision.

In the Color Glass Condensate framework similar
correlations~\cite{Dusling:2009ni,Dusling:2013oia,Dusling:2012wy} and
the mass hierarchy of average $p_T$~\cite{McLerran:2013oju} arise as a
result of gluon saturation in the proton and nuclear
wavefunctions. How much of the observed behavior can be explained as
such an initial state effect\footnote{Here initial state refers to the
  initial state of primary collisions, not to the initial state of
  hydrodynamical evolution.}, and how to differentiate initial state
effects from hydrodynamical final state
effects~\cite{Bzdak:2013zma,Bozek:2013sda} is at the time of this
writing under intense study. If it turns out that the apparently
hydrodynamical behavior in $p+Pb$ collisions can be explained as an
initial state effect, then we may wonder whether the hydrodynamical
behavior in $Pb+Pb$ collisions could be explained as an initial state
effect as well. On the other hand, if this is not possible, and fluid
dynamics is the most viable description of $p+Pb$ collisions, then the
properties of QCD matter are even more surprising than thought so far.

\section{Summary}

Fluid dynamics has been very successful in explaining the features of
bulk, \emph{i.e.} low $p_T$, particle production in ultrarelativistic
heavy-ion collisions. We have seen that in particular anisotropies of
particle production can be explained if the rescatterings among
particles are so frequent that the system is approximately thermal,
that the equation of state of such a matter has many degrees of
freedom and relatively hard, and that the shear viscosity coefficient
over entropy ratio of the produced matter has very low value at some
temperature. Providing further experimental constraints on the
equation of state as well as figuring out the temperature dependence
and the precise value of the minimum of $\eta/s$ will require a lot of
work. The present studies in event-by-event fluctuations and recent
advances in modeling the pre-equilibrium processes in heavy-ion
collisions are very helpful for this goal, but it is not yet clear how
the elliptic flow of photons in $A+A$ collisions and the $p+Pb$
collisions at LHC fit in the overall picture. After a few years' work
we will know.

\section*{Acknowledgements}

This review is based on a talk~\cite{confinement} given in the 10th
Conference on Quark Confinement and the Hadron Spectrum (Confinement
X), Oct 2012, M\"unich, Germany. I thank Hannu Holopainen and Iurii
Karpenko for helpful discussions. This work was supported by BMBF
under contract no.~06FY9092.

Readers may view, browse, and/or download material copyrighted by the
American Physical Society (Figs.~\ref{massorder}, \ref{Paul},
\ref{huichao} and~\ref{photonproblem}) for temporary copying purposes
only, provided these uses are for noncommercial personal
purposes. Except as provided by law, this material may not be further
reproduced, distributed, transmitted, modified, adapted, performed,
displayed, published, or sold in whole or part, without prior written
permission from the American Physical Society.

\end{document}